\documentclass[conference]{IEEEtran}
\IEEEoverridecommandlockouts

\usepackage{tabularx}
\usepackage{booktabs}
\usepackage{threeparttable}
\usepackage{multirow}

\usepackage[ruled,vlined,linesnumbered]{algorithm2e}

\usepackage{subcaption,wrapfig}
\usepackage{graphicx}

\usepackage{amsmath,amsfonts,amssymb,amsthm}
\usepackage{mathtools}

\usepackage{hyperref}

\newtheorem{definition}{\textbf{Definition}}
\newtheorem{theorem}{\textbf{Theorem}}

\usepackage{cite}
\def\BibTeX{{\rm B\kern-.05em{\sc i\kern-.025em b}\kern-.08em
    T\kern-.1667em\lower.7ex\hbox{E}\kern-.125emX}}
\begin{document}

\title{PrivAR: Client-Side Privacy Framework for Real-Time Location-Based Augmented Reality
 }

\author{
\IEEEauthorblockN{
Shafizur Rahman Seeam\textsuperscript{1},
Ye Zheng\textsuperscript{1},
Zhengxiong Li\textsuperscript{2},
Yidan Hu\textsuperscript{1}
}
\IEEEauthorblockA{
\textsuperscript{1}Rochester Institute of Technology, Rochester, NY, USA\\
\textsuperscript{2}University of Colorado Denver, Denver, CO, USA\\
\{ss6365, yz7290, yidan.hu\}@rit.edu, zhengxiong.li@ucdenver.edu
}
}

\maketitle

\begin{abstract}

Location-based augmented reality (LB-AR) applications, such as Pok\'emon Go, rely on sub-second GPS updates to deliver responsive and immersive user experiences. However, this high-frequency location reporting introduces serious privacy risks. Unlike traditional Location-Based Services (LBS), LB-AR demands real-time protection under strict latency and quality-of-service (QoS) constraints, while providing strong per-location and trajectory-level privacy guarantees. Existing privacy mechanisms struggle to satisfy these requirements: they either introduce prohibitive latency, significantly degrade application utility, or fail to defend against trajectory inference attacks.

To address this challenge,  we present PrivAR, the first client-side privacy framework for real-time LB-AR. PrivAR introduces two lightweight mechanisms: (i) Planar Staircase Mechanism (PSM),  which uses a staircase-shaped distribution to generate noisy locations with strong per-location privacy, low expected distortion, and minimal computational overhead; and (ii) Planar Staircase Mechanism with Intermediate (PSM-I), an extension of PSM that generates a device-resident intermediate trajectory and selectively reuses previously perturbed outputs when insufficient drift is observed, thereby strengthening trace-level privacy while preserving high QoS. We provide theoretical analysis, extensive evaluation on two public mobility datasets and our GeoTrace dataset, and validate PrivAR in a Pok\'emon GO–style Android prototype. Results show that PrivAR improves AR QoS (game score) by up to 50\% and increases attacker Bayes risk by up to 1.8$\times$, while incurring only 0.06 ms of per-update overhead (less than 0.2\% of end-to-end latency).

\end{abstract}


\section{Introduction}\label{introduction}
Augmented Reality (AR) overlays virtual content onto the physical world to create interactive real-world experiences. Location-Based Augmented Reality (LB-AR) is a prominent class of AR applications in which a user’s live geographic location is a core input to the system, directly influencing how virtual content is rendered and experienced. Driven by platforms such as ARKit~\cite{AppleArkit} and ARCore~\cite{ARcoreGoogle}, and supported by emerging wearable devices (e.g., AR glasses)~\cite{Hololens}, LB-AR has grown into a multi-billion-dollar ecosystem.

However, the always-on location streams that enable immersive LB-AR experiences also make them uniquely privacy-invasive. For example, Pok\'emon Go can request updates up to 13 times per minute~\cite{kotaku}. Such high-frequency reporting lets service providers (e.g., Niantic, Inc.) accumulate fine-grained mobility traces even within short sessions, from which sensitive information can be inferred, such as home and workplace locations, daily routines, and social interactions. Public scrutiny of LB-AR data practices has intensified~\cite{saudi, backlash}, while regulatory frameworks such as GDPR~\cite{GDPR} impose increasingly strict requirements on the collection and processing of personal location data. As a result, safeguarding real-time location streams in LB-AR has become both a practical systems challenge and a broader societal concern.

\begin{figure}[t]  
    \centering
    \includegraphics[width=0.45\textwidth]{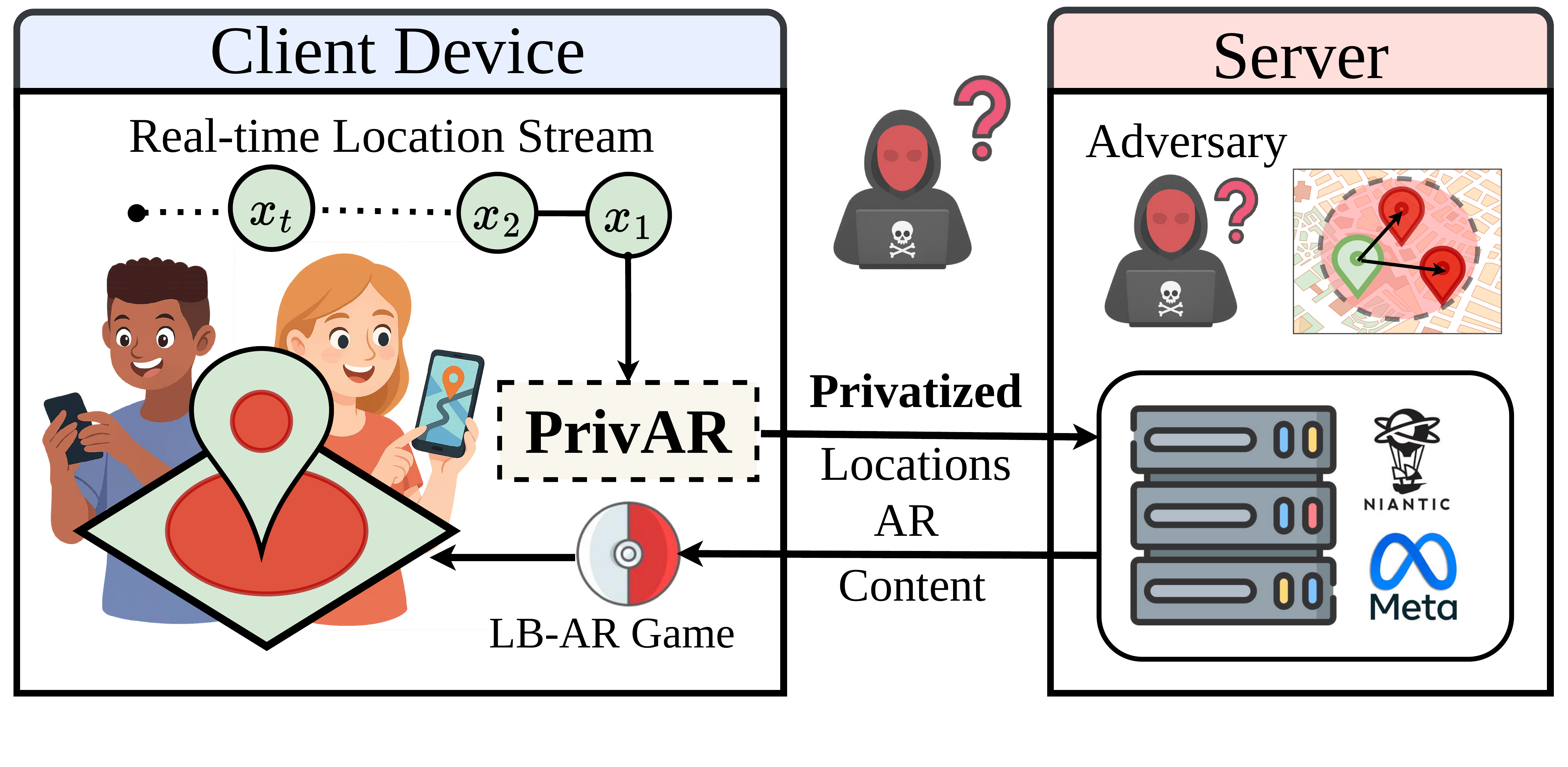} 
    \caption{End-to-end PrivAR-enabled LB-AR system.}
    \label{fig:teaser}
\end{figure}

Privacy protection in LB-AR presents far greater challenges than in traditional Location-Based Services (LBS). Traditional LBS, such as place search (e.g., Yelp~\cite{yelp}) or advertising (e.g., GroundTruth~\cite{groundtruth}), typically issue sparse requests and can often tolerate noticeable latency and coarse perturbation without degrading user experience. In contrast, LB-AR consumes a continuous stream of location updates as part of an interactive rendering and gameplay loop. In this work, we target the common class of LB-AR applications (e.g., location-based games and social AR) whose core interactions tolerate moderate location imprecision (formalized in Sec.~\ref{sec:problem}). Within this scope, any practical client-side protection must satisfy three requirements simultaneously:
\textbf{(R1) Low latency:} incurring sub-millisecond on-device computation per update (and no additional network round trips) to sustain high update rates on resource-constrained hardware;
\textbf{(R2) Per-location and trace-level privacy:} providing strong real-time protection for each released location and for the trajectory under continuous reporting, resisting inference attacks that exploit spatio-temporal correlations; and
\textbf{(R3) High quality of service (QoS):} keeping released privatized locations with sufficiently low expected error to preserve correct AR functionality.
In addition, the mechanism should be \textbf{seamlessly deployable}: it must run entirely on the user device, require no server-side modifications, and integrate as a plug-and-play component in existing LB-AR pipelines, where location streams may be consumed by multiple backend services that are not assumed fully trusted with raw mobility traces.

Unfortunately, mainstream location privacy mechanisms fail to satisfy these requirements at once. Cryptographic approaches (e.g.,~\cite{ShaoFine14}) provide strong privacy but incur heavy computation, violating \textbf{R1}. Spatial cloaking (e.g.,~\cite{gruteser2003anonymous}) is lightweight but offers heuristic protection and remains vulnerable to inference, falling short of \textbf{R2}. Local Differential Privacy mechanisms (e.g.,~\cite{wang2022srr}) provide formal guarantees but inject excessive noise that degrades AR QoS, violating \textbf{R3}. Geo-indistinguishability (GeoInd)~\cite{AndresGeo13} offers a promising alternative by calibrating noise to geographic distance. Among GeoInd mechanisms~\cite{BordenabeOpt14, atmaca2024privacy, biswas2022privic, chatzikokolakis2015constructing}, the Planar Laplace Mechanism (PLM) is appealing from a systems perspective due to its simplicity and efficient on-device execution. However, applying PLM directly to high-frequency LB-AR streams exposes two limitations: (i) substantial per-location distortion due to its inherent radial displacement, degrading QoS; and (ii) rapid privacy degradation and vulnerability to trace inference under continuous reporting.
These limitations underscore the need for GeoInd mechanisms that deliver millisecond-level latency, strong privacy guarantees, and high QoS for real-time LB-AR.


We present PrivAR, a client-side privacy framework  for real-time LB-AR systems. PrivAR runs entirely on the user’s device, requires no changes to the AR server, and integrates as a plug-and-play component in existing LB-AR pipelines (Fig.~\ref{fig:teaser}). PrivAR introduces two lightweight mechanisms: \romannumeral 1) the Planar Staircase Mechanism (PSM) and \romannumeral 2) the Planar Staircase Mechanism with Intermediate (PSM-I). PSM ensures $\epsilon$-GeoInd with improved utility by employing an exponentially decaying, staircase-shaped perturbation profile to strategically inject noise while preserving efficient on-device sampling. Compared to PLM, this design reduces the likelihood of producing excessively distant noisy locations, improving per-update QoS. While PSM improves per-update location quality, it does not address the instability that arises when perturbation is applied independently over long, high-frequency location streams. To address this, PSM-I extends PSM with a device-resident privacy buffer: an intermediate trajectory that is never revealed to the server. The system releases only a filtered public stream derived from this buffer, breaking the direct linkage between consecutive true locations and reported outputs. Building on this buffer, PSM-I applies selective reporting: when the user’s movement is small, the system reuses the most recently released location instead of generating a new noisy report.
We summarize our contributions below:

\begin{itemize}
\item  To the best of our knowledge, this is the first work to systematically study client-side privacy protection for real-time LB-AR systems under practical system constraints.
\item We propose PrivAR, a fully client-side, plug-and-play privacy layer that requires no server-side changes and integrates into existing LB-AR pipelines.
\item We design PSM to improve per-location QoS under GeoInd, and PSM-I,
which builds on PSM, to strengthen trajectory-level protection under
continuous reporting via an intermediate trajectory and selective release.
\item Through extensive evaluation on two public mobility datasets, we show that PSM consistently improves per-location QoS, achieving up to $2\times$ lower error compared to PLM, while providing comparable
practical privacy protection. We further show that PSM-I substantially increases resistance to trace inference attacks, improving Bayes risk against strong trace inference adversaries.
\item We implement an end-to-end Android LB-AR prototype, inspired by Pokémon Go, to demonstrate the practicality of PrivAR in real systems. Experiments on a real-world mobility dataset collected from the prototype (\emph{GeoTrace}) show that PSM improves AR QoS (game score) by up to 50\% compared to PLM, while PSM-I significantly strengthens trace privacy, increasing the Bayes risk of inference attacks by up to $1.8\times$. Across all settings, PrivAR incurs negligible runtime overhead (less than 0.2\%) relative to an unprotected system.
\end{itemize}
The code and data are available in the \href{https://github.com/AnonymousUserGitbhub/PrivAR.git}{PrivAR} repository.

\section{Preliminary}\label{preli}

\subsection{Geo-Indistinguishability}

\begin{definition}[$\epsilon$-Geo-Indistinguishability (GeoInd)~\cite{AndresGeo13}]\label{definitiongeo}
 A randomized mechanism $\mathcal{M}: \mathcal{X} \rightarrow \mathcal{Z}$ satisfies $\epsilon$-GeoInd if for all input locations $x,x' \in \mathcal{X}$,
and all measurable sets $S \subseteq \mathcal{Z}$,
\begin{equation}\label{geo-ind}
\Pr[M(x)\in S] \le e^{\epsilon d(x,x')} \Pr[M(x')\in S].
\end{equation}  
where $d(x,x')$ is the Euclidean distance between $x$ and $x'$.
\end{definition}

\subsection{Planar Laplace Mechanism }\label{plm}
The Planar Laplace Mechanism (PLM)~\cite{AndresGeo13} achieves GeoInd on a continuous 2D plane by making a user’s location $x \in \mathbb{R}^2$ indistinguishable within a radius $r$. It reports a noisy location $z \in \mathbb{R}^2$ drawn from PDF as:

\begin{equation}\label{eq:planarlaplace}
    \mathbf{D}_\epsilon (x)(z) = \frac{\epsilon^2}{2\pi}e^{-\epsilon d(x,z)}
\end{equation}
where $\frac{\epsilon^2}{2\pi}$ is the normalization factor; larger $\epsilon$ weakens privacy.



\subsection{Problem formulation}\label{sec:problem}

\subsubsection{System Model}
We consider real-time location-based augmented reality (LB-AR) applications that
consume location updates to deliver interactive content (e.g., location-based games
such as \textit{Pok\'emon GO}~\cite{PokemonGO} and social platforms such as \textit{Campfire}~\cite{Campfire}).
We target settings where \emph{moderate} location imprecision does not compromise functionality,
and we explicitly exclude AR applications that require centimeter-level geometric registration
or safety-critical localization.
At each time $t\in\{1,\ldots,T\}$, the client device obtains a location fix $x_t\in\mathbb{R}^2$
(e.g., from the device location service using GPS/assisted GPS). The LB-AR client application
requests updates at a high rate (typically every few seconds in common deployments) and
forwards them to server(s) to fetch location-dependent content and execute application logic.
PrivAR is deployed on the client and perturbs each location update before
transmission, requiring no server-side changes.

\subsubsection{Threat model}\label{sec:threat}
We assume the client device is fully trusted and under the user's control, and that it can run PrivAR correctly. We do not consider a compromised client OS or sensor spoofing/jamming attacks; addressing such threats would require additional trusted hardware or OS-level enforcement beyond the scope of this work. The LB-AR application, although developed by the service provider, is assumed to execute only authorized operations as permitted by the user, without performing hidden data collection or unauthorized transmissions. It sends only user-approved location data to the server. In contrast, the remote server, which provides AR content and game logic, is modeled as an \emph{honest-but-curious} adversary: it performs all functionalities correctly but is not trusted to preserve the privacy of user location data. It may accidentally leak location information or deliberately monetize it by sharing with third parties for commercial purposes~\cite{FTC}.
Additionally, we consider a passive network-level eavesdropper capable of observing all outgoing real-time location data transmitted from the client to the server.

\subsubsection{Design goals}

We aim to design PrivAR, a client-side privacy framework for LB-AR that perturbs real-time location streams before transmission, with the following design goals:

\begin{itemize}

  \item \textbf{Rigorous Privacy.}  
  PrivAR must provide privacy protection at single-location and trajectory (trace) levels. Theoretically, it should provide provable privacy guarantees at both levels. In practice, it should defend against two representative attacks: i) \textit{Single-location inference}, inferring a user's location from a single obfuscated report using Bayesian inference~\cite{cherubinFast19}; and ii) \textit{Trace-based inference}, reconstructing user traces from an observed noisy location stream by exploiting spatio-temporal correlations across successive releases~\cite{dong2023location,yu2023privacy}.

  \item \textbf{High QoS.} PrivAR must preserve sufficient location accuracy to support AR interaction. While moderate localization noise is tolerable, it should maintain the expected error within a range that ensures high QoS.

 \item \textbf{Low Latency.} PrivAR must introduce minimal latency during per-location perturbation to ensure it meets the strict timing requirements of LB-AR applications without compromising responsiveness or user experience.

 \item \textbf{Seamless Deployability.} PrivAR must be lightweight and easily integrable into existing LB-AR apps, requiring minimal changes on the client side and \emph{no modifications} to server-side logic or communication protocols.

\end{itemize}

\section{Baseline Solution and Its Limitations}\label{sec:metric:lb-ar}
We use the Planar Laplace Mechanism (PLM) as the natural per-update baseline for LB-AR due to its
simplicity, closed-form sampling, and $\epsilon$-GeoInd guarantee (Sec.~II-B)~\cite{AndresGeo13}.
In this section we focus on (i) how PLM is implemented efficiently on-device and (ii) why naive
independent per-update PLM is insufficient for real-time LB-AR.

\subsubsection{Efficient on-device sampling}
PLM is typically sampled in polar form. Let $u=z-x$, $r=\|u\|$, and $\theta$ be the polar angle.
The induced radius density is
\begin{equation}
p_R(r)=\epsilon^2\,r\,e^{-\epsilon r},\qquad r\ge 0,
\label{eq:plm_radius_pdf}
\end{equation}
and $\theta\sim \mathrm{Unif}(0,2\pi)$. The client samples $r\sim p_R$ and $\theta$ independently and
outputs $z=x+(r\cos\theta,r\sin\theta)$. In our microbenchmarks, PLM runs in sub-millisecond to
millisecond time on representative mobile devices (Table~\ref{tab:runtime}), meeting LB-AR latency
requirements.

\subsubsection{Why per-update PLM is insufficient for LB-AR}
Applying PLM \emph{independently at every high-frequency update} exhibits two fundamental limitations:
(L1) noticeable per-update QoS loss due to typical displacement magnitudes, and (L2) weak trajectory-level
protection under streaming due to composition and temporal correlation.

\textbf{L1: Per-update QoS loss from Jacobian-induced radius mode shift.}
\label{sec:plm:radial}
Although PLM's planar density is maximized at the origin, the radius density in
Eq.~\eqref{eq:plm_radius_pdf} includes the polar-area (Jacobian) factor $r$, shifting the mode away from
$0$ to
\begin{equation}
r^\star=\arg\max_r p_R(r)=1/\epsilon.
\end{equation}
Thus the most likely displacement magnitude is on the order of $1/\epsilon$ (and $\mathbb{E}[R]=2/\epsilon$),
which can be large under privacy-relevant $\epsilon$. For example, at $\epsilon=0.1\,\mathrm{m}^{-1}$, the
radius distribution peaks at $\approx 10$\,m, making moderate-to-large per-update jumps common. In LB-AR,
these jumps manifest as visible jitter and unstable anchoring, degrading interactive QoS (Fig.~\ref{pdfplmpsm}).

\textbf{L2: Weak trajectory-level protection under high-frequency streaming} \label{sec:plm:trace}
\paragraph{Privacy loss under composition}
For a trajectory $\mathbf{x}=\{x_1,\dots,x_T\}$, releasing $\mathcal{M}_\epsilon(x_t)$ independently at each
timestamp causes privacy loss to grow with the number of releases (under sequential composition).
At LB-AR rates (e.g., one fix every 4--5\,s), a 20-minute session yields $T\approx 240$--$300$ releases,
so either per-update $\epsilon$ must be set very small (hurting QoS) or the trajectory-level guarantee
becomes weak.

\paragraph{Inference from correlated motion}
LB-AR motion is temporally correlated: adjacent true locations are close, and movement is smooth. A noisy output at every timestamp provides many correlated observations, enabling denoising and
trajectory reconstruction via filtering/smoothing or learned models~\cite{yu2023privacy}. In addition, LB-AR deployments often involve multiple cloud servers (e.g., gameplay logic and analytics) that consume location updates asynchronously.
As a result, in the absence of an explicit client-side buffering layer, related true
locations can be perturbed and released repeatedly over time and across requests, allowing
adversaries to aggregate multiple noisy views of the same underlying motion even if identical
perturbations are reused within a single timestamp.

These limitations motivate PrivAR's design: we need a per-update mechanism that concentrates more probability mass near the origin (to improve QoS) and a stream-aware design that mitigates trace-level leakage under continuous reporting.

\begin{figure}[t!]  
    \centering
    \includegraphics[width=0.45\textwidth]{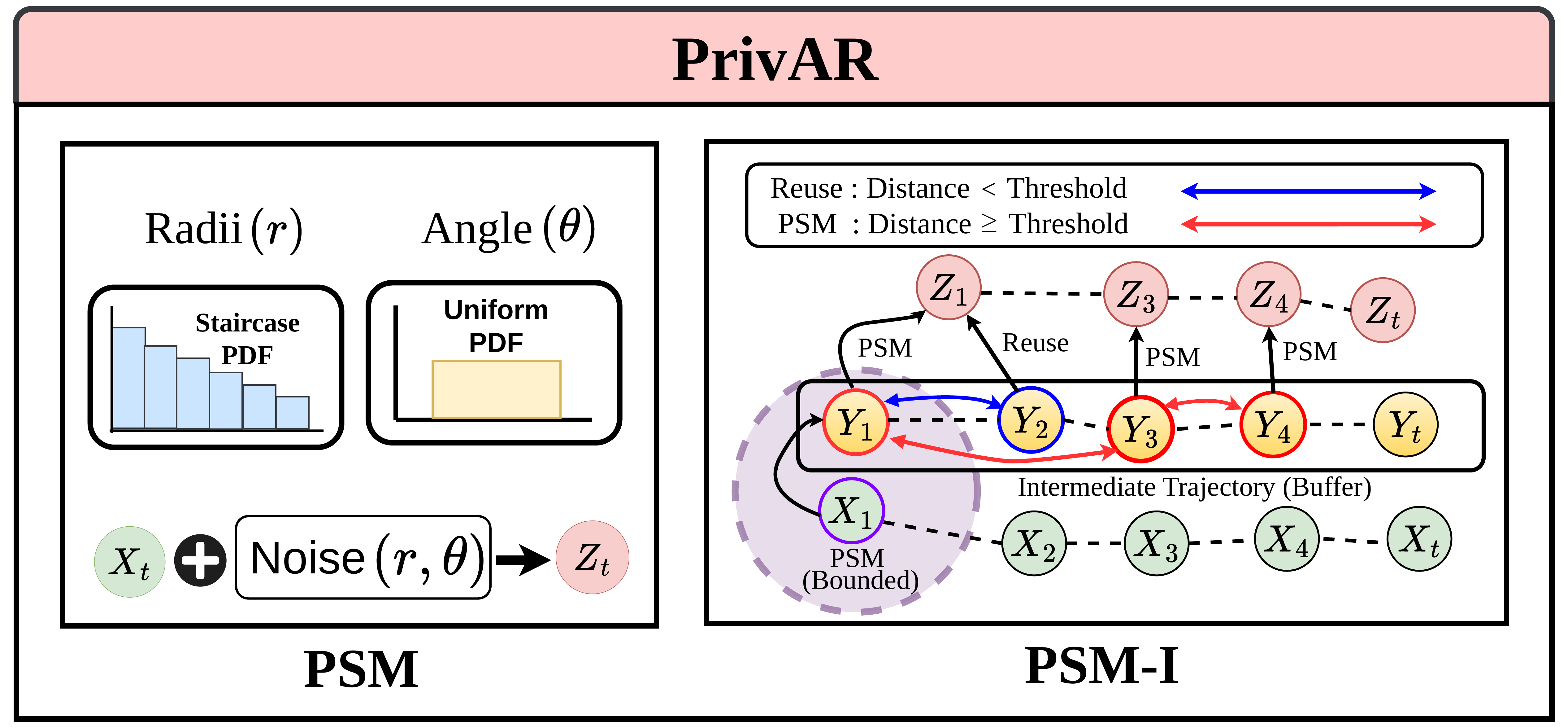} 
    \caption{PrivAR overview. PSM provides per-location GeoInd.
PSM-I uses an on-device intermediate buffer and applies selective
reporting: it applies PSM after displacement above a
threshold and otherwise reuses the last PSM-perturbed release.
} 

    \label{fig:overview}
\end{figure}

\section{Privacy for LB-AR Applications }\label{mechanisms}
This section presents PrivAR, illustrated in Fig.~\ref{fig:overview}, a client-side
privacy framework designed to overcome the limitations of the baseline approach. We first
introduce the Planar Staircase Mechanism (PSM) to address \textbf{L1}. Building on PSM, we then propose the Planar Staircase Mechanism with Intermediate (PSM-I), which further mitigates \textbf{L2}.

\subsection{Planar Staircase Mechanism (PSM)}
\label{sec:psm}

To address PLM’s per-update QoS loss (L1), we design PSM to reshape the displacement distribution so that small radii receive more probability mass while preserving a GeoInd-style guarantee and lightweight on-device sampling.

\subsubsection{Mechanism definition}
PSM partitions the plane around the true location $x\in\mathbb{R}^2$ into concentric annuli of width
$\Delta>0$. Let the $i$-th annulus have inner/outer radii $a_i=(i-1)\Delta$ and $b_i=i\Delta$.
Let $q=e^{-\epsilon\Delta}$.
PSM samples an annulus index $I\in\mathbb{N}^+$ via
\begin{equation}
\label{eq:psm_annulus_prob}
\Pr[I=i]=(1-q)q^{i-1},\qquad i\ge 1,
\end{equation}
then samples a point uniformly over \emph{area} within the selected annulus by drawing
$u\sim\mathrm{Unif}(0,1)$ and $\theta\sim\mathrm{Unif}(0,2\pi)$ and setting
\begin{equation} \label{eq:psm_radius_sample}
r=\sqrt{a_I^2+u(b_I^2-a_I^2)},\qquad z=x+r(\cos\theta,\sin\theta).
\end{equation}
Let $I_x(z):=\left\lceil \|z-x\|/\Delta\right\rceil$ denote the annulus index of $z$ w.r.t.\ $x$.
Since the $i$-th annulus area equals $A_i=\pi(b_i^2-a_i^2)=\pi\Delta^2(2i-1)$, the induced planar
density is piecewise-constant:
\[
p_x(z)=\frac{(1-q)\,q^{I_x(z)-1}}{A_{I_x(z)}}
=\frac{(1-q)\,q^{I_x(z)-1}}{\pi\Delta^2\,(2I_x(z)-1)}.
\]
We summarize the detailed procedure in Algorithm~\ref{alg:psm}.

\begin{algorithm}[t!]
\caption{Planar Staircase Mechanism (PSM)}
\label{alg:psm}
\DontPrintSemicolon
\KwIn{True location $x\in\mathbb{R}^2$, privacy parameter $\epsilon$, step size $\Delta$}
\KwOut{Perturbed location $z\in\mathbb{R}^2$}
$q\leftarrow e^{-\epsilon\Delta}$\;
Sample $I\in\mathbb{N}^+$ with $\Pr[I=i]=(1-q)q^{i-1}$\;
$a_I\leftarrow (I-1)\Delta,\;\; b_I\leftarrow I\Delta$\;
$u\sim \mathrm{Unif}(0,1)$,\;
$r\leftarrow \sqrt{a_I^2+u(b_I^2-a_I^2)}$\;
$\theta\sim \mathrm{Unif}(0,2\pi)$\;
$z\leftarrow x+r(\cos\theta,\sin\theta)$\;
\Return $z$
\end{algorithm}

\subsubsection{Why PSM places more mass near the origin than PLM}
PSM admits simple closed-form comparisons against PLM.

\textbf{(i) Larger near-origin mass.}
Let $R=\|z-x\|$ be the displacement radius. Under PSM,
\begin{equation}
\Pr[R\le \Delta]=\Pr[I=1]=1-q=1-e^{-\epsilon\Delta}.
\label{eq:psm_first_ring_mass}
\end{equation}
Under PLM, the radius CDF is $F_{\mathrm{PLM}}(r)=1-e^{-\epsilon r}(1+\epsilon r)$, hence
\begin{equation}
\Pr_{\mathrm{PLM}}[R\le \Delta]=1-e^{-\epsilon\Delta}(1+\epsilon\Delta).
\label{eq:plm_first_mass}
\end{equation}
Since $(1+\epsilon\Delta)>1$, Eq.~\eqref{eq:psm_first_ring_mass} is strictly larger than
Eq.~\eqref{eq:plm_first_mass} for any $\epsilon,\Delta>0$.
For small $\epsilon\Delta$, $\Pr_{\mathrm{PSM}}[R\le\Delta]\approx \epsilon\Delta$ while
$\Pr_{\mathrm{PLM}}[R\le\Delta]\approx (\epsilon\Delta)^2/2$, showing that PSM can allocate
orders-of-magnitude more probability to very small displacements.

\begin{figure}[!t]
    \centering
    \includegraphics[width=0.50\linewidth]{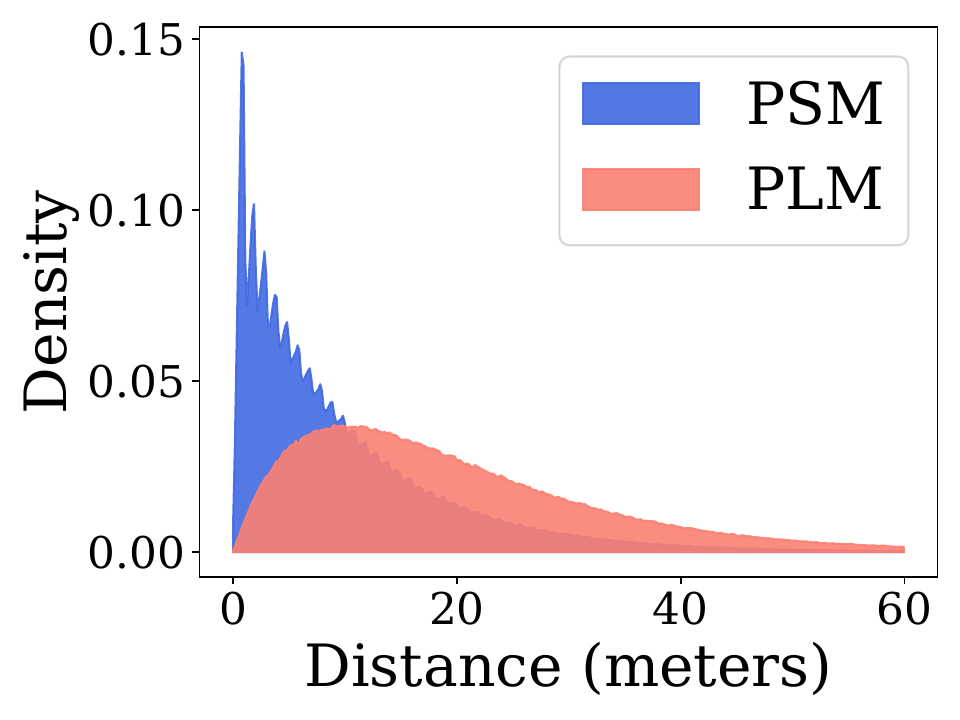}
    \caption{Radial displacement density at $\epsilon=0.1$: PSM concentrates mass near the origin, whereas PLM peaks around $1/\epsilon$.}

    \label{pdfplmpsm}
\end{figure}

\textbf{(ii) Lighter large-error tails.}
At annulus boundaries $r=i\Delta$, PSM has a geometric tail:
\[
\Pr[R>i\Delta]=\Pr[I>i]=q^{i}=e^{-\epsilon i\Delta}.
\]
In contrast, PLM has
\[
\Pr_{\mathrm{PLM}}[R>i\Delta]=e^{-\epsilon i\Delta}(1+\epsilon i\Delta)\;\ge\;e^{-\epsilon i\Delta},
\]
which includes an additional polynomial factor $(1+\epsilon r)$, making large-radius outliers more
likely. 
Table~\ref{tab:comparison_table} and Fig.~\ref{pdfplmpsm} confirm that PSM reduces displacement and large-error outliers using $10^5$ i.i.d.\ radius samples at $\epsilon=0.1$ with $\Delta=1$.

\subsubsection{Privacy guarantee (GeoInd-style)}
Because PSM samples uniformly over \emph{area} within each annulus, its planar density includes the
factor $(2I_x(z)-1)^{-1}$ and therefore does not decrease by exactly $e^{-\epsilon\Delta}$ between
adjacent annuli. Nevertheless, PSM admits a GeoInd-style bound under the $\Delta$-quantized metric
$d_\Delta(x,x'):=\Delta\lceil d(x,x')/\Delta\rceil$.

\begin{theorem}
\label{thm:PSM}
Let $d_\Delta(x,x')=\Delta\lceil d(x,x')/\Delta\rceil$ and
$k=\lceil d(x,x')/\Delta\rceil \ge 1$.
PSM satisfies $\tilde\epsilon$-GeoInd under $d_\Delta$:
for all $x,x'\in\mathbb{R}^2$ and measurable $S$,
\[
\Pr[\mathrm{PSM}(x)\in S]\le
e^{\tilde\epsilon d_\Delta(x,x')}
\Pr[\mathrm{PSM}(x')\in S],
\]
where $\tilde\epsilon=\epsilon+\ln(3)/\Delta$.
\end{theorem}

\begin{proof}[Proof sketch]
From the closed-form density,
$\frac{p_x(z)}{p_{x'}(z)}=
q^{I_x(z)-I_{x'}(z)}\frac{2I_{x'}(z)-1}{2I_x(z)-1}$.
Since $|I_x(z)-I_{x'}(z)|\le k$ and $q=e^{-\epsilon\Delta}$, the first factor is at most
$e^{\epsilon d_\Delta(x,x')}$, and
$I_{x'}(z)\le I_x(z)+k$ implies
$\frac{2I_{x'}(z)-1}{2I_x(z)-1}\le 2k+1$.
Thus
$p_x(z)\le(2k+1)e^{\epsilon d_\Delta(x,x')}p_{x'}(z)$.
Using $2k+1\le3^k$ and $d_\Delta=\Delta k$ yields
$(2k+1)e^{\epsilon d_\Delta(x,x')}
\le
e^{(\epsilon+\ln(3)/\Delta)d_\Delta(x,x')}$.
Integrating over any measurable set $S$ completes the proof.
\end{proof}


\vspace{0.25em}
\noindent\textbf{Remark.}
The additive term $\ln 3 / \Delta$ reflects a worst-case slack arising from
uniform-over-area sampling and the use of coarse upper bounds in the analysis.
This bound is conservative: increasing $\Delta$ tightens $\tilde{\epsilon}$
toward $\epsilon$, and in practice the effective privacy loss is significantly
smaller. Our evaluation shows that PSM achieves empirical Bayes-risk privacy comparable
to PLM at the same $\epsilon$, even with $\Delta=1$, while substantially
improving QoS (Sec.~\ref{privacysimulations}).


\begin{table}[!t]
\centering
\caption{Empirical displacement error under PLM and PSM.}
\label{tab:comparison_table}
\begin{tabular}{ccc}
\toprule
Mechanism & Mean Distance (m) & Max Distance (m)  \\
\midrule
PLM         & 20.001 & 155.524 \\
PSM  & \textbf{10.044} & \textbf{141.719}  \\

\bottomrule
\end{tabular}
\end{table}

\subsubsection{Runtime overhead}
PLM sampling can be implemented via inverse transform sampling, which evaluates the Lambert-$W$ function (the $-1$ branch)~\cite{AndresGeo13}, whereas PSM sampling uses only random variates and arithmetic (geometric/uniform draws and a square root).
In a repeated single-sample microbenchmark at $\epsilon=0.1$ (200{,}000 iterations), we measure $5.19\times 10^{-6}$~s/sample for PLM and $4.27\times 10^{-6}$~s/sample for PSM on our machine, indicating that both mechanisms are suitable for on-device sampling.




\textbf{Bounded PSM ($F_{r_b}$).}
We extend PSM with a bounded variant $F_{r_b}(\cdot)$ that enforces a hard cap on
the displacement radius, guaranteeing $d(x,z)\le r_b$.
Let $m=\lceil r_b/\Delta\rceil$ and $q=e^{-\epsilon\Delta}$; for simplicity, we
set $r_b=m\Delta$.
Bounded PSM samples a ring index $I\in\{1,\ldots,m\}$ from the truncated geometric
distribution
$
\Pr[I=i\mid I\le m]=\frac{(1-q)q^{i-1}}{1-q^m},
$
then samples $(r,\theta)$ as in Eq.~(\ref{eq:psm_radius_sample}) within the annulus
$[(I-1)\Delta,I\Delta]$ and outputs $z=x+r(\cos\theta,\sin\theta)$.
Compared to bounded PLM (i.e., planar Laplace conditioned on $R\le r_b$), bounded
PSM concentrates more probability mass near the origin
(e.g., $\Pr[R\le\Delta]=\frac{1-q}{1-q^m}$), improving AR stability while
eliminating outliers beyond $r_b$.
We next characterize the privacy guarantees of bounded PSM.

\begin{theorem}[Bounded PSM]
\label{thm:bPSM}
Let $r_b=m\Delta$, $q=e^{-\epsilon\Delta}$, and
$d_\Delta(x,x')=\Delta\lceil d(x,x')/\Delta\rceil$.
Bounded PSM $F_{r_b}$ satisfies $(\tilde\epsilon,\delta_b)$-Geo-Indistinguishability
under $d_\Delta$: for any $x,x'$ with $k=\lceil d(x,x')/\Delta\rceil<m$ and any
measurable set $S$,
\[
\Pr[F_{r_b}(x)\in S]
\le
e^{\tilde\epsilon d_\Delta(x,x')}
\Pr[F_{r_b}(x')\in S]
+\delta_b,
\]

where $\tilde\epsilon=\epsilon+\ln(3)/\Delta$ and
$\delta_b=\dfrac{q^{m-k}-q^m}{1-q^m}$.
\end{theorem}

\begin{proof}[Proof sketch]
Let $p_x^{(b)}(z)$ denote the density of $F_{r_b}(x)$.
For outputs $z$ whose ring index satisfies $I_x(z)\le m-k$, truncation does not
affect either input, and the same ring-index argument as in
Theorem~\ref{thm:PSM} yields
$p_x^{(b)}(z)\le (2k+1)e^{\epsilon d_\Delta(x,x')}p_{x'}^{(b)}(z)$.
Using $2k+1\le 3^k$ and $d_\Delta=\Delta k$, we obtain
$p_x^{(b)}(z)\le e^{\tilde\epsilon d_\Delta(x,x')}p_{x'}^{(b)}(z)$.
The remaining probability mass lies in the boundary rings $I>m-k$, which may be
present for one input but truncated for the other; its total probability is
bounded by $\delta_b$.
Integrating over $S$ completes the proof.
\end{proof}

\subsection{Planar Staircase Mechanism with Intermediate (PSM-I)}
\label{sec:psmi}
\subsubsection{Overview}
PSM improves per-update QoS, but applying any per-location mechanism independently at high frequency still
enables trace inference and incurs rapid privacy loss
under composition as described in \textbf{L2}. 
We thus propose PSM-I to address \textbf{L2} in high-frequency LB-AR systems. 
PSM-I reduces this exposure through two complementary ideas:
(i) an on-device \emph{intermediate privacy buffer} that reduces direct linkage between the released stream and the true stream, and (ii) a \emph{selective refresh} policy that releases a fresh privatized
location only when the buffered state has moved sufficiently far.

\paragraph{Intermediate privacy buffer} 
PSM-I runs on the client and maintains an intermediate sequence
$\{y_t\}_{t=1}^{T}$ that is never revealed to the server. Public releases are generated from $\{y_t\}$ rather than directly from $\{x_t\}$. While the intermediate buffer can in principle be generated using richer motion models or side information, we adopt this displacement-copying update because it requires no additional background knowledge (e.g., road maps, map matching, or learned priors) and preserves short-term motion with minimal overhead. Specifically, at session start, the client samples an initial intermediate point $y_1$ near the initial true location $x_1$ using a bounded PSM to limit initial utility loss, and then releases $z_1$ by perturbing $y_1$ using standard PSM. 
For each subsequent timestamp $t\ge 2$, 
PSM-I propagates the intermediate state by applying the observed displacement of the true trajectory (from $x_{t-1}$ to $x_t$) to the previous intermediate point $y_{t-1}$. This creates a coherent buffered path that preserves short-term motion needed for LB-AR QoS, while ensuring that public releases are anchored to the buffered
state rather than directly to the true locations.
As a result, the adversary is limited to inferring a coarsened, reused trajectory, reducing temporal resolution.

\paragraph{Selective refresh and reuse}
LB-AR applications often request location updates at a high rate even when the user moves only slightly between consecutive updates. Releasing a fresh perturbed location at every timestamp provides little additional benefit to the application in these cases, but it gives the server many correlated observations that can be fused to strengthen trace inference. PSM-I therefore refreshes only when the buffered state moves beyond a threshold $\delta$; otherwise it reuses the previous released point. 
Selective reuse induces a many-to-one mapping from intermediate locations to released outputs: multiple consecutive intermediate locations may correspond to the same released point (e.g., $z_t=z_{t+1}=\cdots$ while $y_t$ continues to evolve). Consequently, an unchanged release does
not reveal whether the user remained stationary or moved within the session-specific threshold $\delta$, weakening
temporal linkability and hindering trace-level inference.

\begin{table}[t!]
\centering
\renewcommand{\arraystretch}{1.2}
\begin{tabular}{|c|c|}
\hline
$a$ &
$\sin^2(\Delta \phi / 2)
+ \cos(\phi_{x_{t-1}})\cdot \cos(\phi_{x_t})
  \cdot \sin^2(\Delta \lambda / 2)$ \\ 
\hline
$b$ &
$\sin(\Delta \lambda)\cdot \cos(\phi_{x_t})$ \\ 
\hline
$c$ &
$\cos(\phi_{x_{t-1}})\cdot \sin(\phi_{x_t})
 - \sin(\phi_{x_{t-1}})\cdot \cos(\phi_{x_t})
   \cdot \cos(\Delta \lambda)$ \\ 
\hline
$d$ &
$\sin(\theta)\cdot \sin(d_h / R_E)\cdot \cos(\phi_{y_{t-1}})$ \\ 
\hline
$e$ &
$\cos(d_h / R_E) - \sin(\phi_{y_{t-1}})\cdot \sin(\phi_{y_t})$ \\ 
\hline
$f$ &
$\cos(\phi_{y_{t-1}})\cdot \sin(d_h / R_E)\cdot \cos(\theta)$ \\ 
\hline
$g$ &
$\sin(\phi_{y_{t-1}})\cdot \cos(d_h / R_E)$ \\ 
\hline
\end{tabular}
\caption{Expressions used in
Eq.~\ref{eq:haversinedistance},
Eq.~\ref{eq:direction}, and
Eq.~\ref{eq:intermediatelocation}. Here,
$\Delta\phi = \phi_{x_t}-\phi_{x_{t-1}}$ and
$\Delta\lambda = \lambda_{x_t}-\lambda_{x_{t-1}}$.}
\label{tab:equationsconstant}
\end{table}

\subsubsection{Detailed design}
PSM-I maps a stream of true locations $\{x_t\}_{t=1}^{T}$ to released locations $\{z_t\}_{t=1}^{T}$, with all computation performed locally on the client device.
All locations are represented in latitude-longitude form: for any location $v$ we write $v=(\phi_v,\lambda_v)$. 
\emph{In practice, PSM-I operates on latitude–longitude coordinates but assumes
a local planar approximation, under which Euclidean distance is used for privacy
accounting and theoretical analysis.}
We denote one invocation of PSM by $\mathcal{F}(\cdot)$ with privacy
parameter $\epsilon$, and a bounded PSM by $\mathcal{F}_{r_b}(\cdot)$ that truncates the sampled radius to ensure outputs lie within great-circle distance $r_b$ of the input.
Internally, the client maintains: (i) the current intermediate state $y_t$;
(ii) a session-specific refresh threshold $\delta$; and (iii) a reference pair
$(y_{\mathrm{ref}}, z_{\mathrm{ref}})$, where $y_{\mathrm{ref}}$ is the intermediate state at which
the most recent refresh occurred and $z_{\mathrm{ref}}$ is the corresponding released point.

\textit{Initialization ($t=1$)}
Given location $x_1$, the client generates:
\begin{equation}
y_1 \leftarrow \mathcal{F}_{r_b}(x_1), \qquad
z_1 \leftarrow \mathcal{F}(y_1).
\end{equation}

The client then initializes the reference pair $(y_{\mathrm{ref}}, z_{\mathrm{ref}})\leftarrow (y_1, z_1)$.

\subsubsection{Operation ($t\ge 2$)}
For each new true location $x_t$, the client first computes the great-circle distance and initial
bearing from $x_{t-1}$ to $x_t$ (as in~\cite{maria2020measure}):
\begin{equation}\label{eq:haversinedistance}
d_h = \mathrm{Dis}(x_{t-1},x_t)=2 R_E \arcsin\big(\sqrt{a}\big),
\end{equation}
\begin{equation}\label{eq:direction}
\theta=\mathrm{Dir}(x_{t-1},x_t) = \operatorname{atan2}(b,c),
\end{equation}
where $a,b,c$ are given in Table~\ref{tab:equationsconstant}.
PSM-I then applies the same displacement $(d_h,\theta)$ to the previous intermediate point
$y_{t-1}$ using the destination-point great-circle transform~\cite{dis}:
\begin{equation}\label{eq:intermediatelocation}
y_t =\Big(\arcsin(f+g),\; \lambda_{y_{t-1}}+\operatorname{atan2}(d,e)\Big),
\end{equation}
where $d,e,f,g$ are given in Table~\ref{tab:equationsconstant}.

Finally, PSM-I determines whether to refresh the public output by comparing the current intermediate
state to the last refreshed reference:
$d^{(y)}_t \leftarrow \mathrm{Dis}(y_{\mathrm{ref}}, y_t)$.
If $d^{(y)}_t < \delta$, the client reuses the previous release by setting
$z_t \leftarrow z_{\mathrm{ref}}$. Otherwise, it refreshes the release by perturbing the current
intermediate state, $z_t \leftarrow \mathcal{F}(y_t)$, and updates the reference pair as
$(y_{\mathrm{ref}}, z_{\mathrm{ref}})\leftarrow (y_t, z_t)$.
The complete procedure
is summarized in Algorithm~\ref{alg:psmi}.

\begin{algorithm}[t!]
\caption{PSM-I}
\label{alg:psmi}
\DontPrintSemicolon
\LinesNumbered
\SetKwInput{KwIn}{Input}
\SetKwInput{KwOut}{Output}

\KwIn{True stream $(x_1,\ldots,x_T)$; per-refresh budget $\epsilon$; threshold $\delta$; bounding radius $r_b$}
\KwOut{Released stream $(z_1,\ldots,z_T)$}

\BlankLine
\textbf{Initialization ($t=1$)}\;
$y_1 \leftarrow \mathcal{F}_{r_b}(x_1)$\tcp*{bounded PSM}
$z_1 \leftarrow \mathcal{F}(y_1)$\tcp*{first public release}
$(y_{\mathrm{ref}}, z_{\mathrm{ref}}) \leftarrow (y_1, z_1)$\tcp*{reference for reuse}
\BlankLine

\For{$t \leftarrow 2$ \KwTo $T$}{
    $d_h \leftarrow \mathrm{Dis}(x_{t-1}, x_t)$\tcp*{Eq.~\ref{eq:haversinedistance}}
    $\theta \leftarrow \mathrm{Dir}(x_{t-1}, x_t)$\tcp*{Eq.~\ref{eq:direction}}
    $y_t \leftarrow \mathrm{Move}(y_{t-1}, d_h, \theta)$\tcp*{Eq.~\ref{eq:intermediatelocation}}
    
    $d^{(y)}_t \leftarrow \mathrm{Dis}(y_{\mathrm{ref}}, y_t)$\;
    \uIf{$d^{(y)}_t < \delta$}{
        $z_t \leftarrow z_{\mathrm{ref}}$\tcp*{reuse}
    }\Else{
        $z_t \leftarrow \mathcal{F}(y_t)$\tcp*{refresh via PSM}
        $(y_{\mathrm{ref}}, z_{\mathrm{ref}}) \leftarrow (y_t, z_t)$\tcp*{new reference}
    }
}
\Return $(z_1,\ldots,z_T)$\;
\end{algorithm}

\subsubsection{Theoretical analysis}
At $t=1$, PSM-I initializes the intermediate state using bounded PSM and releases a privatized location via a single application of PSM. The privacy guarantee at $t=1$ follows directly from the bounded PSM guarantee in Theorem~\ref{thm:bPSM} and the post-processing property of differential privacy~\cite{DworkAlg14}, yielding $(\tilde{\epsilon},\delta_b)$-GeoInd. For  $t\geq2$, we have the following theorem.
\begin{theorem}[Post-Initialization Trace Privacy]
\label{thm:post-init-privacy}
Let $\mathbf x=(x_1,\ldots,x_T)$ and $\mathbf x'=(x_1',\ldots,x_T')$ be two input
traces, and let $\mathbf y(\mathbf x)$ and $\mathbf y(\mathbf x')$ be the
corresponding intermediate traces generated by PSM-I.
Conditioned on the same initial intermediate state $y_1$, for any measurable
$S\subseteq(\mathbb R^2)^{T-1}$,
\[
\begin{split}
&\Pr[\mathrm{PSM\text{-}I}(\mathbf x)_{2:T}\in S|y_1]
\le\\
&\exp\!\Big(
\tilde\epsilon\sum_{t=2}^T d_\Delta\big(y_t(\mathbf x),y_t(\mathbf x')\big)
\Big)
\Pr[\mathrm{PSM\text{-}I}(\mathbf x')_{2:T}\in S|y_1],
\end{split}
\]
where $\tilde\epsilon=\epsilon+\ln(3)/\Delta$.
\end{theorem}

\begin{proof}[Proof sketch]
From Theorem~\ref{thm:PSM}, PSM satisfies $\tilde\epsilon$-GeoInd.
Let $\mathcal{R}\subseteq\{2,\ldots,T\}$ denote the timestamps at which
PSM-I refreshes (rather than reuses the previous release).
Conditioned on $y_1$, the only randomized outputs are the fresh releases
$\{F(y_t)\}_{t\in\mathcal{R}}$; all reuse steps are deterministic functions
of previously released values and are therefore post-processing.
By sequential composition of GeoInd over the $|\mathcal{R}|\le T-1$
invocations of $F$, the likelihood ratio is bounded by the product of
per-refresh bounds, yielding the stated exponent.
\end{proof}

\section{Simulation Studies}\label{evaluation}
We first evaluate the proposed solutions in terms of QoS, privacy, and latency using public datasets.

\subsection{Experimental Setting} 
\subsubsection{Environment} QoS and privacy evaluations were run on an M2 MacBook Pro (16 GB RAM), while latency was measured across three Android phones (low-end, mid-range, and high-end). All results are averaged over 100 trials.

\subsubsection{Datasets}
We evaluate our methods on two real-world datasets: Geolife~\cite{zheng2008understanding} and T-Drive~\cite{zheng2011t-drive}. We use sub-sampled versions of Geolife and T-Drive. Geolife contains 1.05 million points covering 23,061 km, with a median step of 8 m and 2 s intervals. T-Drive includes 809K points across 4,343 km, with a 49 m median step and 181 s intervals.

\subsubsection{Preprocessing}  
Following prior work~\cite{cherubinFast19,shafizurPicasso2024}, we consider a \(6 \times 6\)~km\(^2\) geographic region discretized into a \(200 \times 200\) grid. Each true location \(x_t = (\phi_t, \lambda_t)\), where \(\phi_t\) and \(\lambda_t\) denote latitude and longitude, is mapped to a corresponding grid cell \(s_t \in \mathcal{S}\). Independently, \(x_t\) is perturbed by the privacy mechanism to produce a noisy observation \(z_t\). The resulting labeled pairs \((z_t, s_t)\) are used to train the attack model.

\subsubsection{Attack model}  
We consider the \emph{trace-based inference attack}~\cite{shafizurPicasso2024}, in which the adversary observes a sequence of perturbed locations \(\{z_1, z_2, \dots, z_T\}\) and exploits spatio-temporal correlations to infer the user's true trace. Following~\cite{cherubinFast19, shafizurPicasso2024}, we adopt the \(k\)-Nearest Neighbors (\(k\)-NN) as the inference model, setting \(k = \log(T)\). This choice is motivated by the universal consistency of \(k\)-NN, which guarantees convergence as sample size \(T \to \infty\)~\cite{stone1977consistent}. We also evaluate an HMM-based attacker that models discretized true locations $\{s_t\}$ as latent states with a first-order Markov mobility prior learned from training traces, and released locations $\{z_t\}$ as noisy observations. Given an observation window of length $L$, the attacker applies Bayesian filtering to estimate $P(s_t \mid z_{t-L+1:t})$. Attack performance is measured by empirical Bayes risk, averaged over traces and time steps.

\subsubsection{Evaluation Metrics} We evaluate mechanisms along three dimensions: QoS, privacy, and latency.

\begin{table*}[t!]
  \centering
  \caption{MNE of PSM-I under varying refresh thresholds $\delta$ for $\epsilon\in\{0.1, 0.5\}$ on Geolife and T-Drive.}
  \label{tab:mne-trpsm}
  \begin{tabular}{l rrrrrr | rrrrrr}
    \toprule
    & \multicolumn{6}{c|}{$\epsilon=0.1$} & \multicolumn{6}{c}{$\epsilon=0.5$} \\
    \cmidrule(lr){2-7} \cmidrule(lr){8-13}
    Dataset 
      & $\delta=3$ & $\delta=5$ & $\delta=10$ & $\delta=20$ & $\delta=50$ & $\delta=100$
      & $\delta=3$ & $\delta=5$ & $\delta=10$ & $\delta=20$ & $\delta=50$ & $\delta=100$ \\
    \midrule
    Geolife 
      & 14.06 & 14.16 & 15.98 & 21.75 & 45.92 & 78.86 
      & 3.60  & 4.81  &  8.86  & 17.94 & 45.90 & 79.06 \\
    T-Drive  
      & 13.90 & 14.13 & 15.54 & 20.09 & 41.40 & 66.69 
      & 3.54  & 4.62  & 8.26  & 16.41 & 40.92 & 66.60 \\
    \bottomrule
  \end{tabular}
  \label{deltatable}
\end{table*}

\paragraph{Quality of Service (QoS)}  
We quantify spatial distortion using Mean Normalized Error (MNE)~\cite{Zhang23Tra}, which measures the average distance between true and perturbed locations across all traces. Let \(\mathcal{T}\) denote the set of traces, with each trace \(\tau_j = \{(x_t^{(j)}, z_t^{(j)})\}_{t=1}^{T_j}\). MNE is defined as:
\begin{equation}
    \mathrm{MNE} = \frac{1}{|\mathcal{T}|} \sum_{j=1}^{|\mathcal{T}|} \frac{1}{T_j} \sum_{t=1}^{T_j} d(x_t^{(j)}, z_t^{(j)})
\end{equation}
where \(d(\cdot, \cdot)\) denotes the Euclidean distance.  Lower MNE implies better QoS by indicating smaller spatial error~\cite{biswas2022privic}.


\paragraph{Privacy}  
We estimate the adversary’s success using empirical Bayes risk~\cite{romanelli2020estimating}. For each attacker (e.g., \(k\)-NN or HMM), let \(f_T\) denote the learned inference procedure that outputs a posterior distribution over grid cells. Given samples \(\{(z_t, s_t)\}_{t=1}^{T}\), where \(z_t\) is the perturbed location and \(s_t\) the true grid cell, the estimated Bayes risk is:
\[
\hat{\mathcal{R}}_{f_T} := 1 - \sum_{z \in \mathcal{Z}} \max_{s \in \mathcal{S}} \Pr(z, s)
\]
where \(\Pr(z, s) = \Pr(z \mid s)\pi(s)\) denotes the joint distribution. Higher \(\hat{\mathcal{R}}_{f_T}\) indicates stronger privacy, as it reflects greater uncertainty in the adversary’s inference.

\paragraph{Latency}
We measure latency as the average time required to perturb a single location update. Lower latency indicates better suitability for real-time LB-AR applications.

\subsection{Simulation results}

\begin{figure}[t!]
  \centering
  \begin{minipage}{0.48\linewidth}
    \centering
    \includegraphics[width=\linewidth]{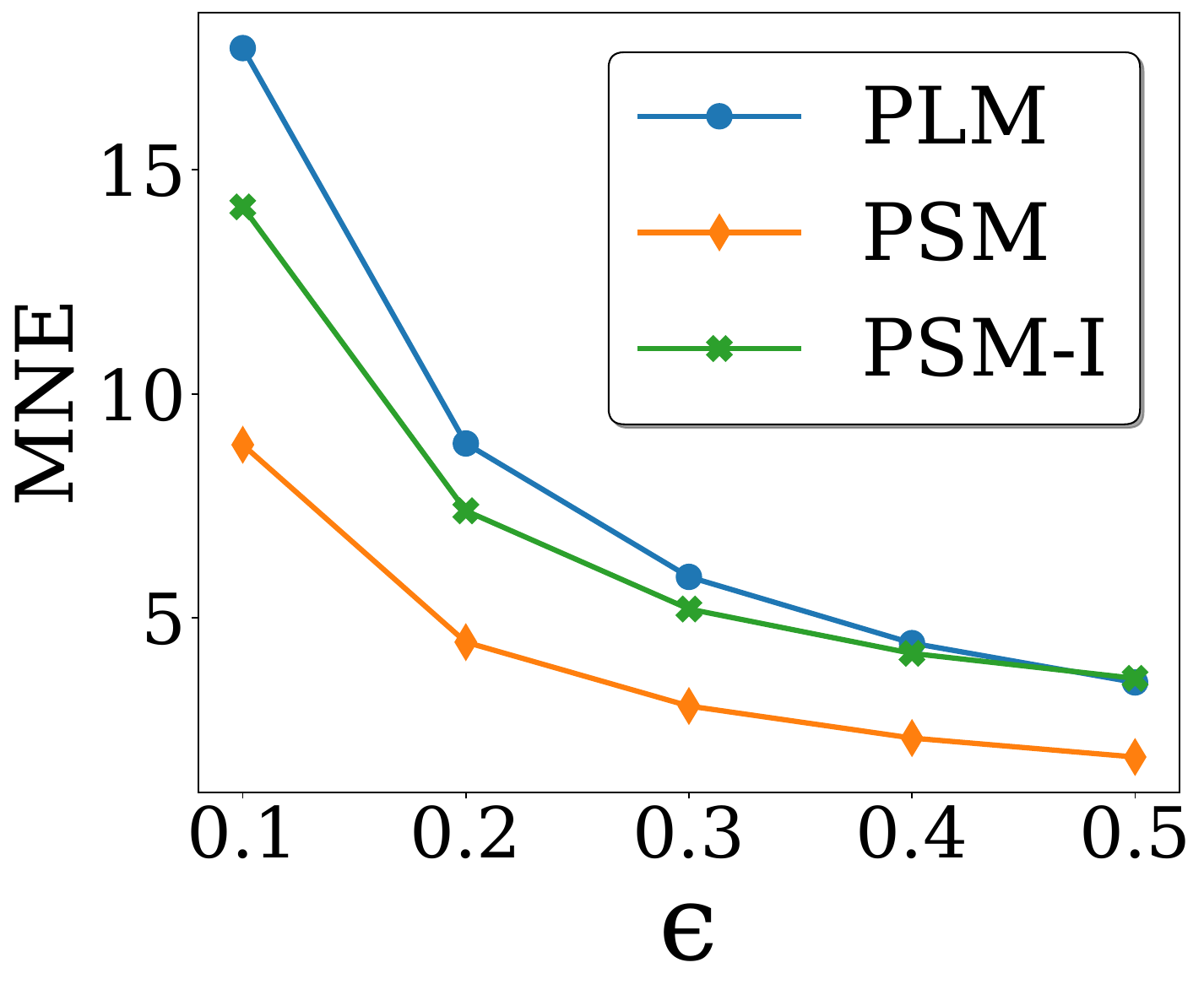}   
    \subcaption{Geolife}
  \end{minipage}\hfill
  \begin{minipage}{0.48\linewidth}
    \centering
    \includegraphics[width=\linewidth]{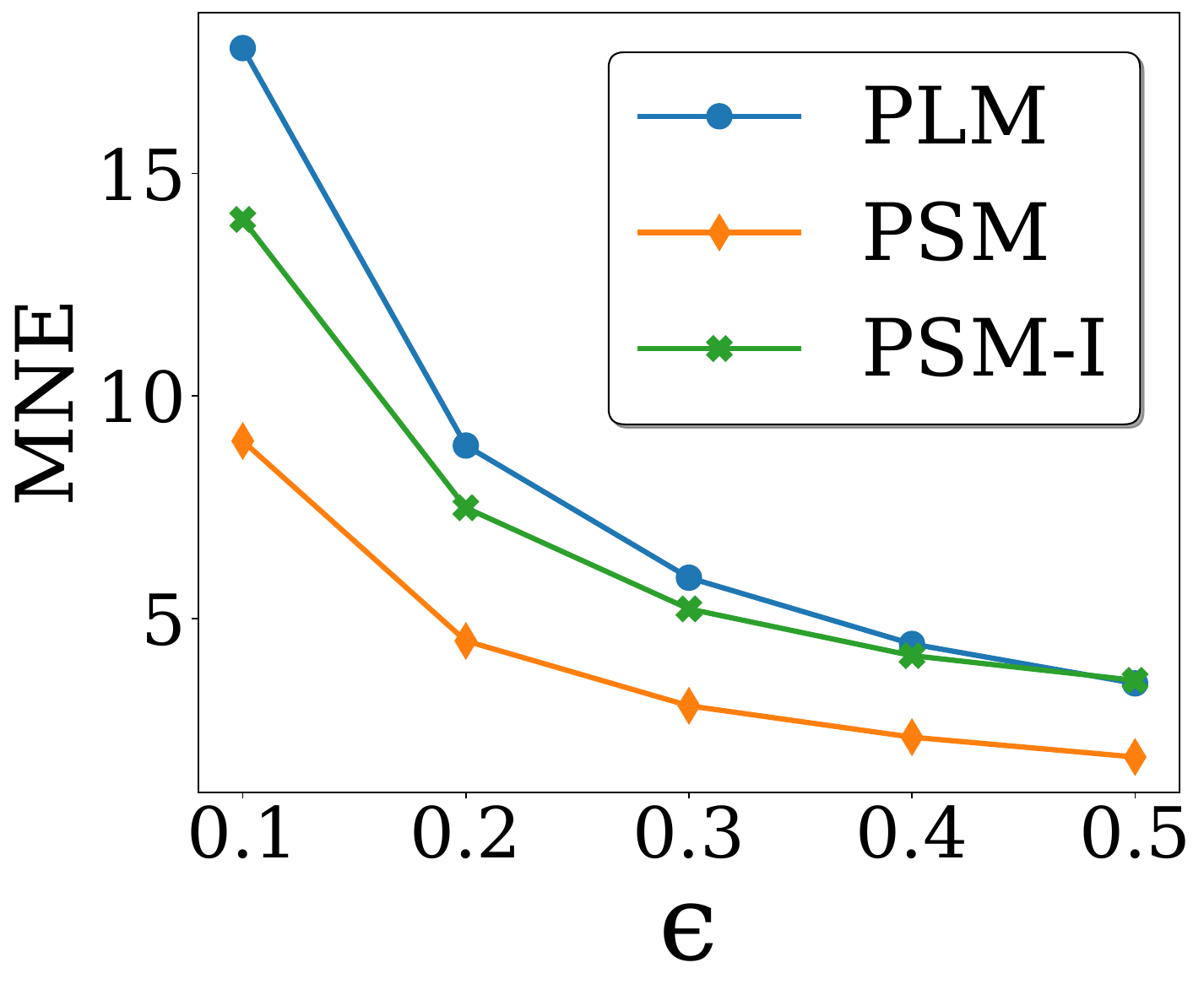}     
    \subcaption{T-Drive}
  \end{minipage}

  \caption{QoS on Geolife and T-Drive: MNE vs. $\epsilon$ (0.1–0.5) for PLM, PSM, PSM-I; lower is better.}
  \label{mne_realworld}
\end{figure}

\subsubsection{QoS}\label{utilityexperiemnts}
Fig.~\ref{mne_realworld} reports the MNE of PLM, PSM, and PSM-I across two datasets, with \(\epsilon\) ranging from 0.1 to 0.5. As expected, MNE decreases with increasing \(\epsilon\) for all mechanisms, since higher privacy budgets introduce less noise, producing perturbed locations closer to the ground truth. PSM consistently achieves the lowest MNE, demonstrating superior QoS, especially under strong privacy settings (i.e., low \(\epsilon\)). In contrast, PLM yields the highest MNE, particularly at lower privacy budgets. While PSM-I incurs slightly higher MNE than PSM, it still significantly outperforms PLM. More importantly, this modest QoS loss is offset by the substantial privacy gains PSM-I provides (see~\ref{privacysimulations}). Table~\ref{deltatable} further shows that increasing the reporting threshold \(\delta\) leads to higher MNE, reflecting the expected QoS–privacy trade-off.

\begin{figure}[t!]
  \centering
  \begin{minipage}{0.48\linewidth}
    \centering
    \includegraphics[width=\linewidth]{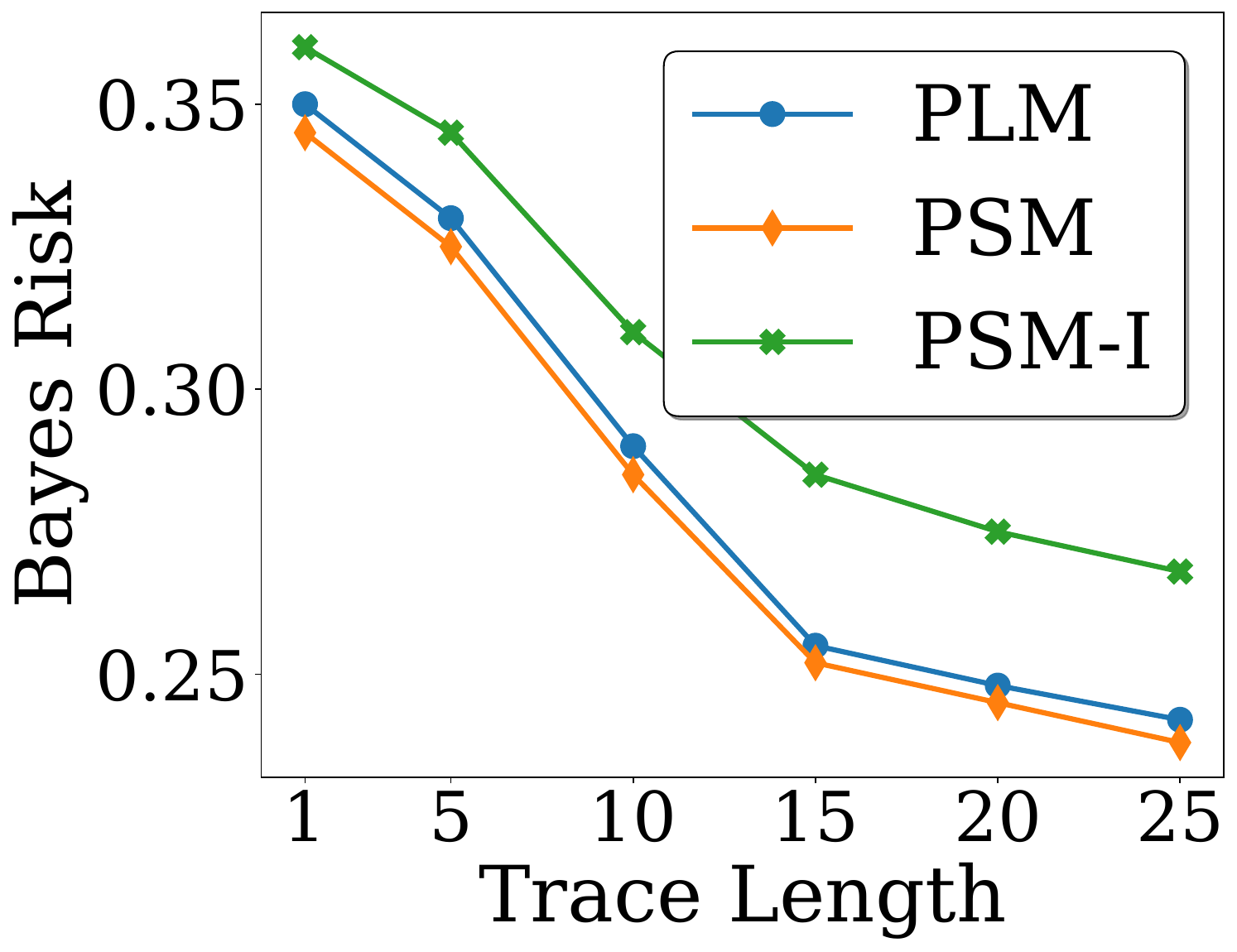}   
    \subcaption{Geolife}
    \label{bayes_geolife}
  \end{minipage}\hfill
  \begin{minipage}{0.48\linewidth}
    \centering
    \includegraphics[width=\linewidth]{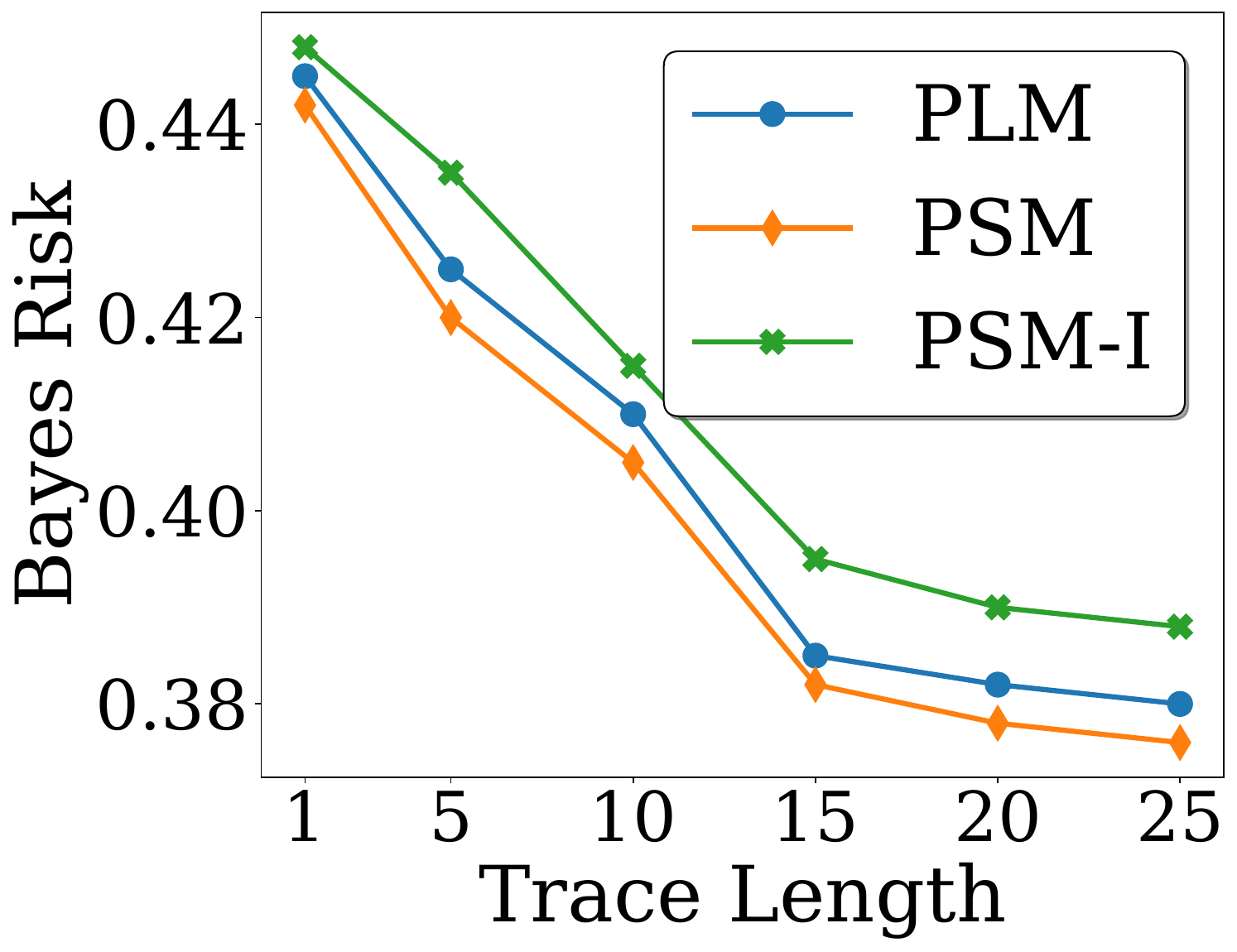}   
    \subcaption{T-Drive}
    \label{bayes_tdrive}
  \end{minipage}

  \caption{Trace-level privacy on Geolife and T-Drive: Bayes risk vs. trajectory length at $\epsilon=0.1$; higher is better.}
  \label{bayes_trace_all}
\end{figure}

\subsubsection{Privacy}\label{privacysimulations}
Figure~\ref{bayes_trace_all} shows the Bayes risk of PLM, PSM, and PSM-I under a \(k\)-NN inference attack at \(\epsilon=0.1\), as trace length increases from 1 to 25. A trace length of 1 corresponds to a single-location attack. As expected, Bayes risk decreases with longer traces as the adversary accumulates spatio-temporal evidence. PSM-I consistently achieves higher Bayes risk than PLM and PSM across both datasets. This improvement stems from two design choices: (i) public releases are generated from a private intermediate trajectory rather than directly from true locations, and (ii) selective refresh induces many-to-one mappings, causing multiple consecutive locations to share the same released point. Together, these effects reduce temporal resolution and limit the adversary’s ability to exploit sequential correlations. The benefit is most pronounced in Geolife (Fig.~\ref{bayes_geolife}), where dense and highly correlated movements (median inter-fix distance \(\approx 8\)~m) enable rapid privacy degradation under PLM and PSM. In contrast, PSM-I preserves substantially higher Bayes risk by reusing prior outputs. In the sparser T-Drive dataset (median inter-fix distance \(=49\)~m; Fig.~\ref{bayes_tdrive}), all mechanisms degrade more gradually, as weaker spatial continuity inherently constrains the attacker’s inference power.

\subsubsection{Latency}\label{sec:latency}

\begin{table}[t!]
\centering
\caption{Per-location runtime (ms) on mobile devices.}
\label{tab:runtime}
\begin{tabular}{lccc}
\toprule
\textbf{Device} & \textbf{PLM} & \textbf{PSM} & \textbf{PSM-I} \\
\midrule
Galaxy A04 (Low-end)     & 0.556 & 0.551 & 0.864 \\
Pixel 6a (Mid-range)     & 0.114       & 0.112       & 0.142 \\
Galaxy S22 Ultra (High-end)    & 0.046       & 0.048       & 0.059 \\
\bottomrule
\end{tabular}
\end{table}

Table~\ref{tab:runtime} reports the per-location perturbation runtime across three Android devices. PLM and PSM exhibit nearly identical runtimes, as both apply single-shot perturbation. PSM-I introduces modest overhead due to its conditional logic, but remains highly efficient, achieving sub-millisecond latency on all tested devices. Even on the low-end Galaxy A04, PSM-I completes perturbation within 0.9~ms, while on modern devices such as the Pixel 6a and Galaxy S22 Ultra, runtimes remain well below 0.2~ms. These latencies are orders of magnitude lower than the 16.7~ms display refresh interval of 60~Hz screens~\cite{deber2015much}, ensuring suitability for latency-sensitive AR applications.

\section{PrivAR: Implementation and Evaluation}

\begin{figure*}[t]
  \centering

  \begin{minipage}[t]{0.71\linewidth}
    \centering
    \includegraphics[width=\linewidth, height=3.5cm]{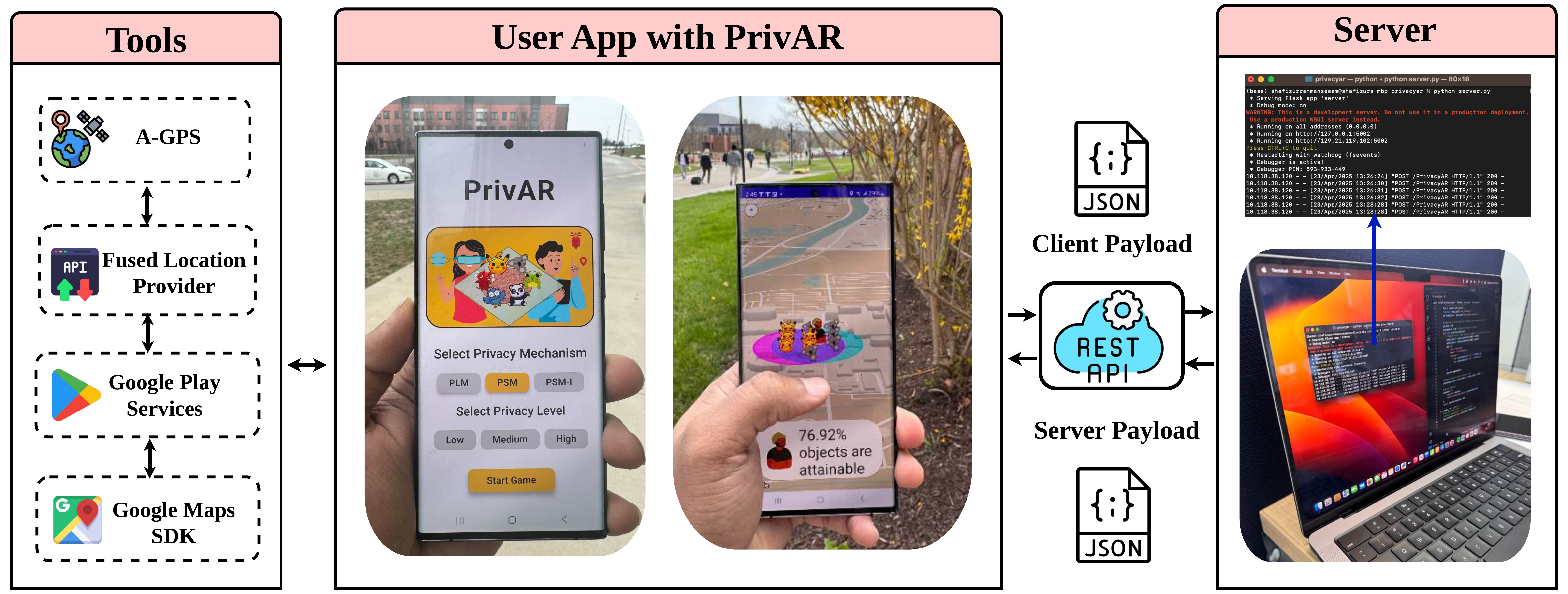}
    
    \caption{In our PrivAR-enabled app, the device perturbs the user’s real-time location before transmitting it to the game server. The server computes nearby virtual objects (e.g., PokéStops) from noisy locations and returns their coordinates in a compact JSON payload, which the client renders locally for a low-latency AR experience.}
    \label{endtoend}
  \end{minipage}%
  \hfill
  \begin{minipage}[t]{0.27\linewidth}
     \includegraphics[width=\linewidth, height=3.65cm]{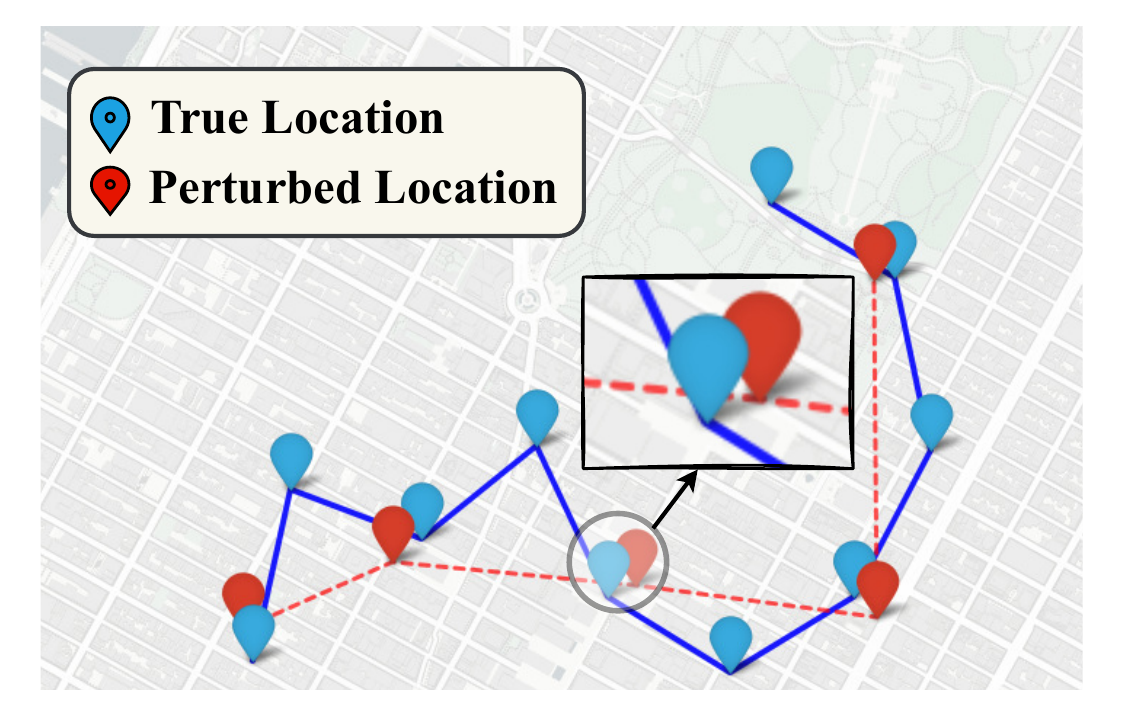}
    \caption{PSM-I releases a new noisy location when movement exceeds a threshold; otherwise, the previous release is reused.}
    \label{trpsmout}
  \end{minipage}%
  
\end{figure*}

This section presents the end-to-end implementation of PrivAR in a custom-built, Pokémon GO–inspired LB-AR application. Figures~\ref{endtoend} and~\ref{trpsmout} illustrate the complete end-to-end workflow and the selective-release behavior of PSM-I, respectively.

\subsection{End-to-End Implementation of LB-AR App}\label{testbed}

\subsubsection{Client-Side Workflow}

The client-side app was developed in Kotlin 2.0.21 using Android Studio Meerkat (2024.3.1) with Android Gradle Plugin 8.9.0. It targets Android 7.0 (API level 24) or higher and is compiled against SDK 35 with Java 11 compatibility.  The app leverages the Fused Location Provider API (Play Services Maps v19.1.0) to capture precise locations, which are perturbed in real time by PrivAR before server transmission. PrivAR is implemented as a self-contained Kotlin plugin and is designed for open-source release as a lightweight, ARCore-compatible library. Its core mechanisms, PSM and PSM-I, were initially prototyped in Python (NumPy 1.21.5, SciPy 1.7.3, Pandas 1.4.2) and reimplemented in Kotlin for efficient on-device deployment.

\paragraph*{App Interface}  The app provides separate configuration and gameplay interfaces. In the configuration UI, users select a privacy mechanism (PLM, PSM, or PSM-I) and a privacy level corresponding to \(\epsilon \in \{0.5, 0.2, 0.1\}\), representing low, medium, and high privacy, respectively. During gameplay, the app visualizes two concentric 100-meter visibility regions centered at the true and perturbed locations. Their intersection determines where virtual objects are rendered. A live counter reports the fraction of objects visible under privacy.

\subsubsection{Server-Side Workflow} The backend server is a lightweight AR content generator implemented in Flask v3.1. Hosted on a local machine (e.g., MacBook Pro), it exposes a RESTful API and communicates with the mobile client (e.g., Samsung Galaxy S22 Ultra) over Wi-Fi. The server listens for HTTP requests at the \texttt{/PrivAR} endpoint, parses incoming JSON payloads, extracts the perturbed location, and dynamically generates nearby virtual game objects.

\subsubsection{Client–Server Workflow}

Gameplay begins when the user selects privacy preferences—e.g., \texttt{PSM} with \texttt{High} via the configuration UI. Upon tapping  \texttt{Start}, the app continuously acquires real-time locations using the Fused Location Provider API,  perturbs them using the selected mechanism, and transmits the perturbed coordinates to the server.\footnote{True locations are shown locally only for evaluation and are never transmitted in real deployments.} The server generates nearby virtual objects based solely on the perturbed input and returns them to the client, which renders only those falling within the overlapping visibility regions.

\paragraph*{Integration in Commercial Apps}  
PrivAR is designed as a drop-in, client-side library that requires no changes to server logic or communication protocols. This design enables straightforward integration into existing LB-AR applications with minimal engineering effort.

\subsection{Practicality of PrivAR}\label{sec:feasibility}

We evaluate the practicality of PrivAR through real-world experiments, focusing on AR-specific QoS, robustness against location inference, and end-to-end latency in a client–server LB-AR workflow.

\subsubsection{Experimental Setup}
Five participants (3 male, 2 female) were each provided with a preconfigured Samsung Galaxy S22 Ultra running our PrivAR-enabled prototype. Participants completed multiple sessions of varying durations, freely moving within designated areas while interacting with the app. During each session, the client received JSON payloads from the server to render virtual objects based on perturbed locations, while logging QoS metrics and ground-truth mobility traces for offline analysis. In total, we collected approximately 127~km of trajectory data across diverse mobility modes, including walking, running, biking, and driving, forming our \emph{GeoTrace} dataset. The anonymized dataset is publicly available in the \href{https://github.com/AnonymousUserGitbhub/PrivAR.git}{PrivAR} repository. On the server side, virtual objects were uniformly generated around perturbed locations under sparse (100~m) and dense (25~m) layouts.

\subsubsection{Evaluation Metrics}

In typical AR games, users capture virtual objects via throwing or proximity-based interactions. We simplify gameplay by assuming that users automatically capture all virtual objects within their visibility region. AR QoS (game score) is evaluated using two metrics:

(i) \emph{Catchable Objects}: the percentage of virtual objects still reachable after perturbation:
\[
\mathcal{C}_{\text{obj}} = \frac{|\mathcal{V} \cap \hat{\mathcal{V}}|}{|\mathcal{V}|} \times 100\%
\]
where \( \mathcal{V} \) and \( \hat{\mathcal{V}} \) denote object sets around the true and perturbed locations.

(ii) \emph{Accumulated Loss}: the total number of objects missed over a session:
\[
\mathcal{L}_{\text{acc}} = \sum_{t=1}^{T} \left( |\mathcal{V}^{(t)}| - |\mathcal{V}^{(t)} \cap \hat{\mathcal{V}}^{(t)}| \right)
\]
where \( T \) is session length and \( \mathcal{V}^{(t)} \) is object set at time \( t \).

\begin{figure}[t]
  \centering
  \begin{subfigure}[t]{0.48\columnwidth}
    \centering
    \includegraphics[width=\linewidth]{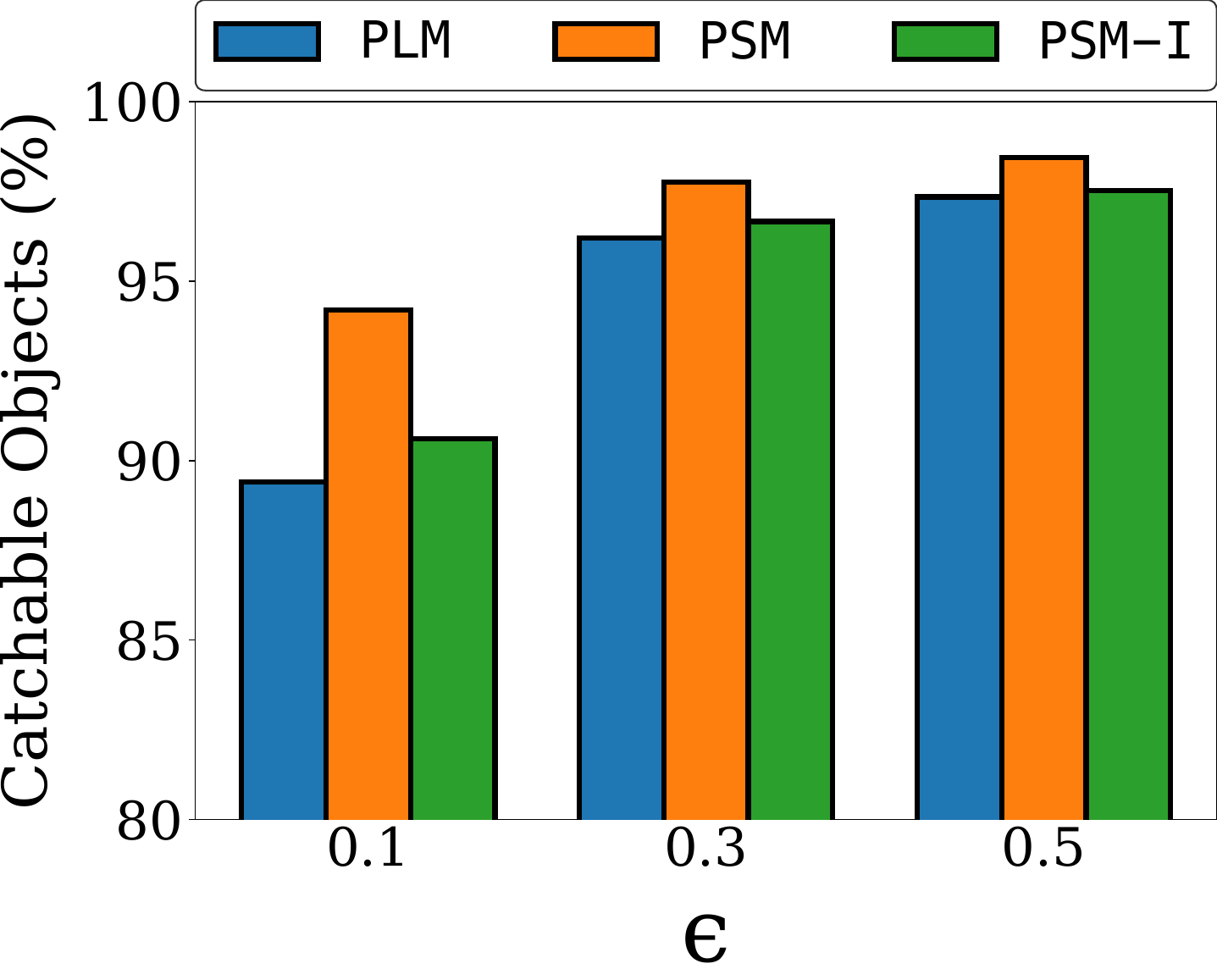}
    \subcaption{Dense Virtual Objects}
    \label{subfig:catchable_objects_a}
  \end{subfigure}
  \hfill
  \begin{subfigure}[t]{0.48\columnwidth}
    \centering
    \includegraphics[width=\linewidth]{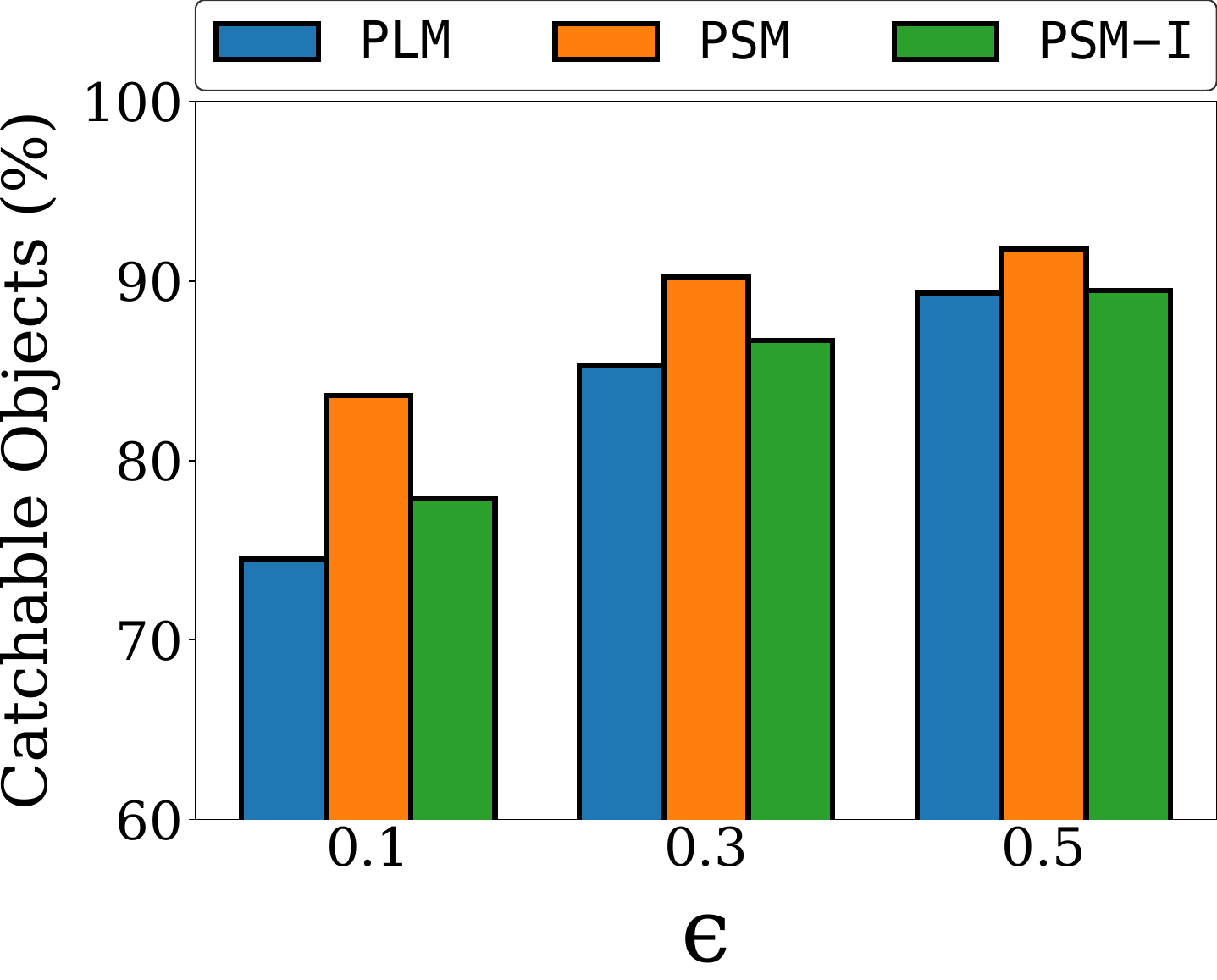}
    \subcaption{Sparse Virtual Objects}
    \label{subfig:catchable_objects_b}
  \end{subfigure}

  \caption{\% Catchable objects vs. $\epsilon$ (0.1--0.5) under (a) dense and (b) sparse virtual-object layouts; higher is better.}
  \label{fig:catchable_objects}
\end{figure}

\begin{figure}[t]
  \centering
  \begin{subfigure}[t]{0.48\columnwidth}
    \centering
    \includegraphics[width=\linewidth]{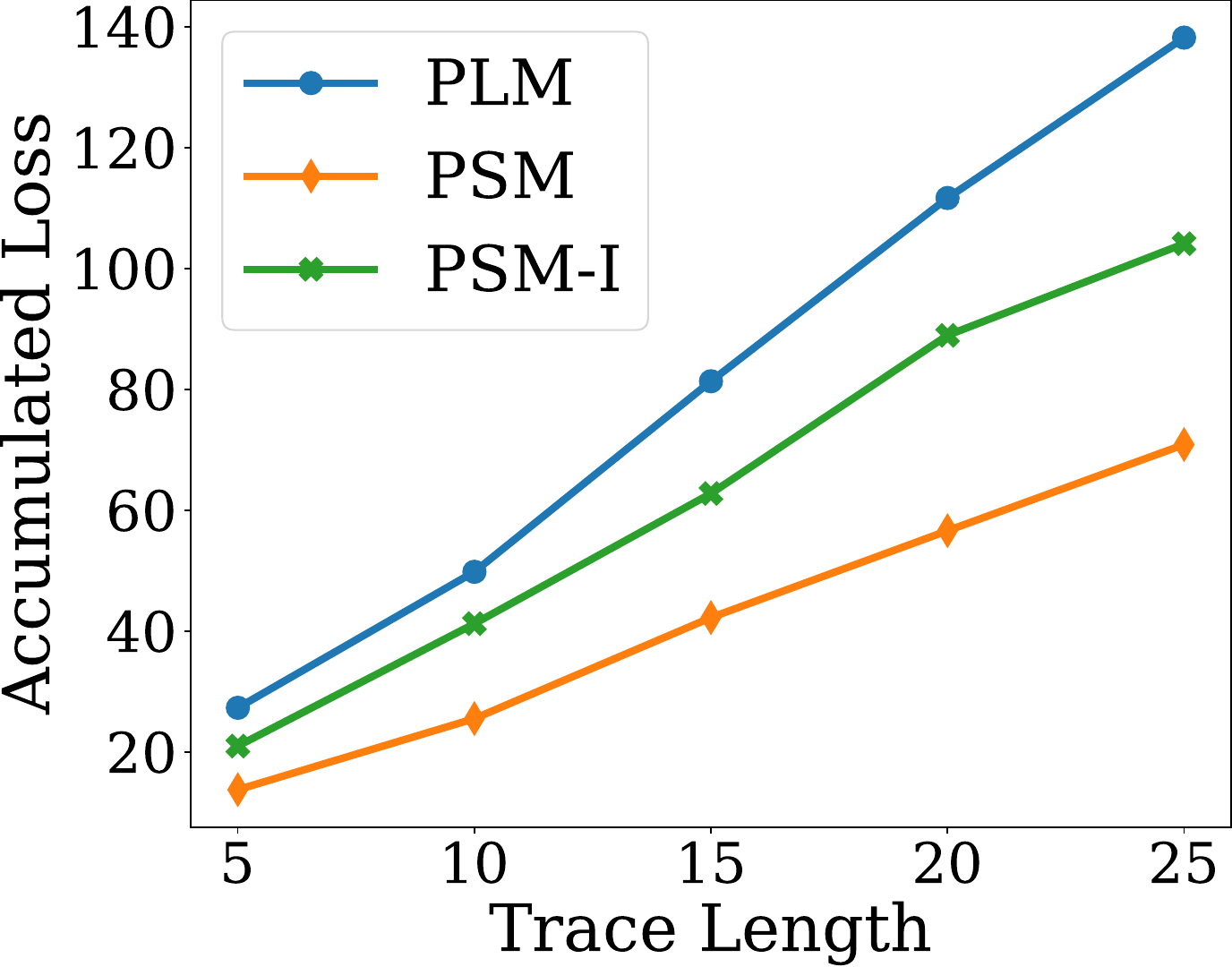}
    \subcaption{Dense Virtual Objects}
    \label{subfig:Accumulated_loss_A}
  \end{subfigure}
  \hfill
  \begin{subfigure}[t]{0.48\columnwidth}
    \centering
    \includegraphics[width=\linewidth]{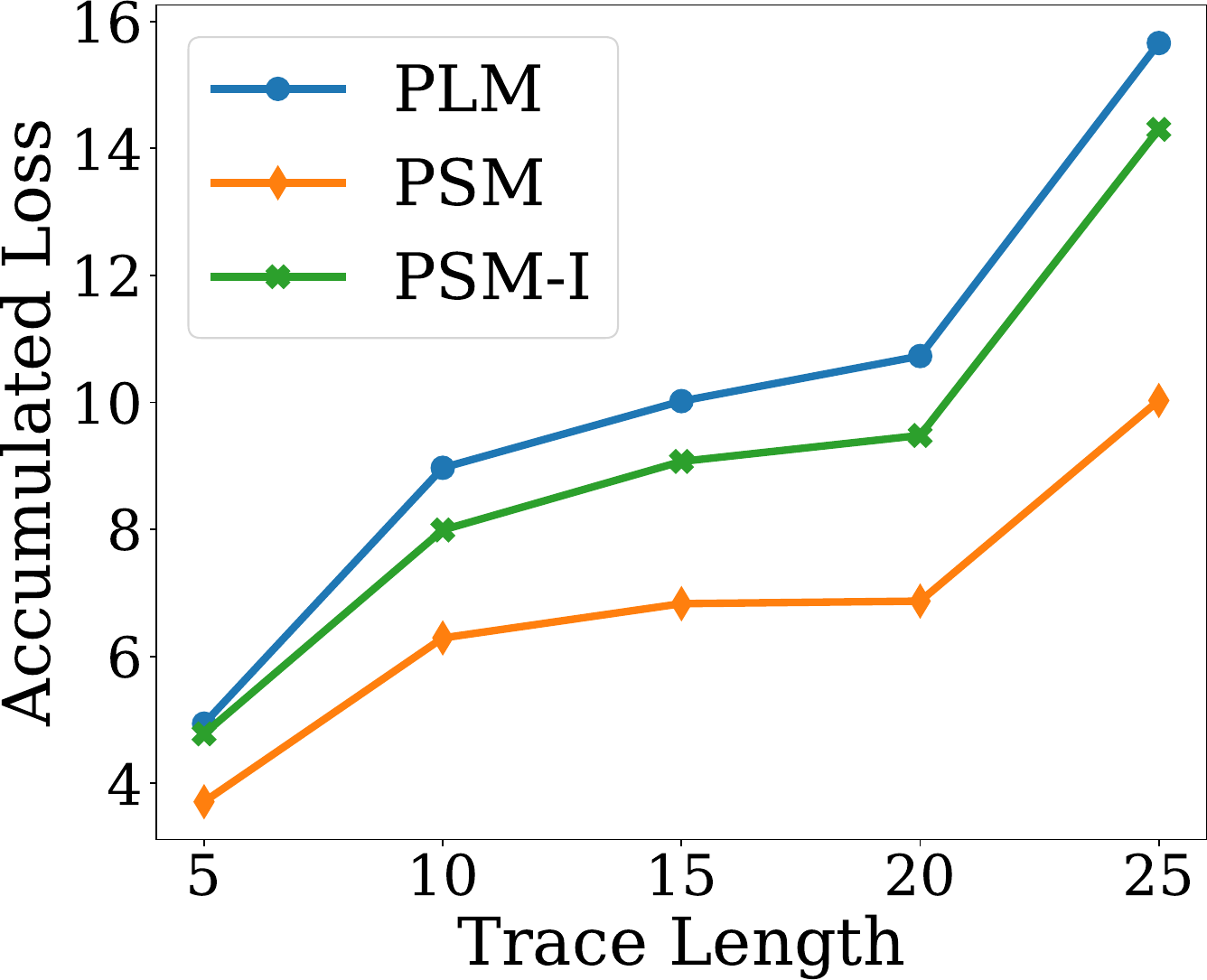}
    \subcaption{Sparse Virtual Objects}
    \label{subfig:Accumulated_loss_B}
  \end{subfigure}

  \caption{Accumulated loss vs. trace length when $\epsilon=0.1$ under (a) dense and (b) sparse virtual-object layouts; lower is better.}
  \label{fig:Accumulated_loss}
\end{figure}

\begin{figure}[t]
    \centering
      \includegraphics[width=4.5cm]{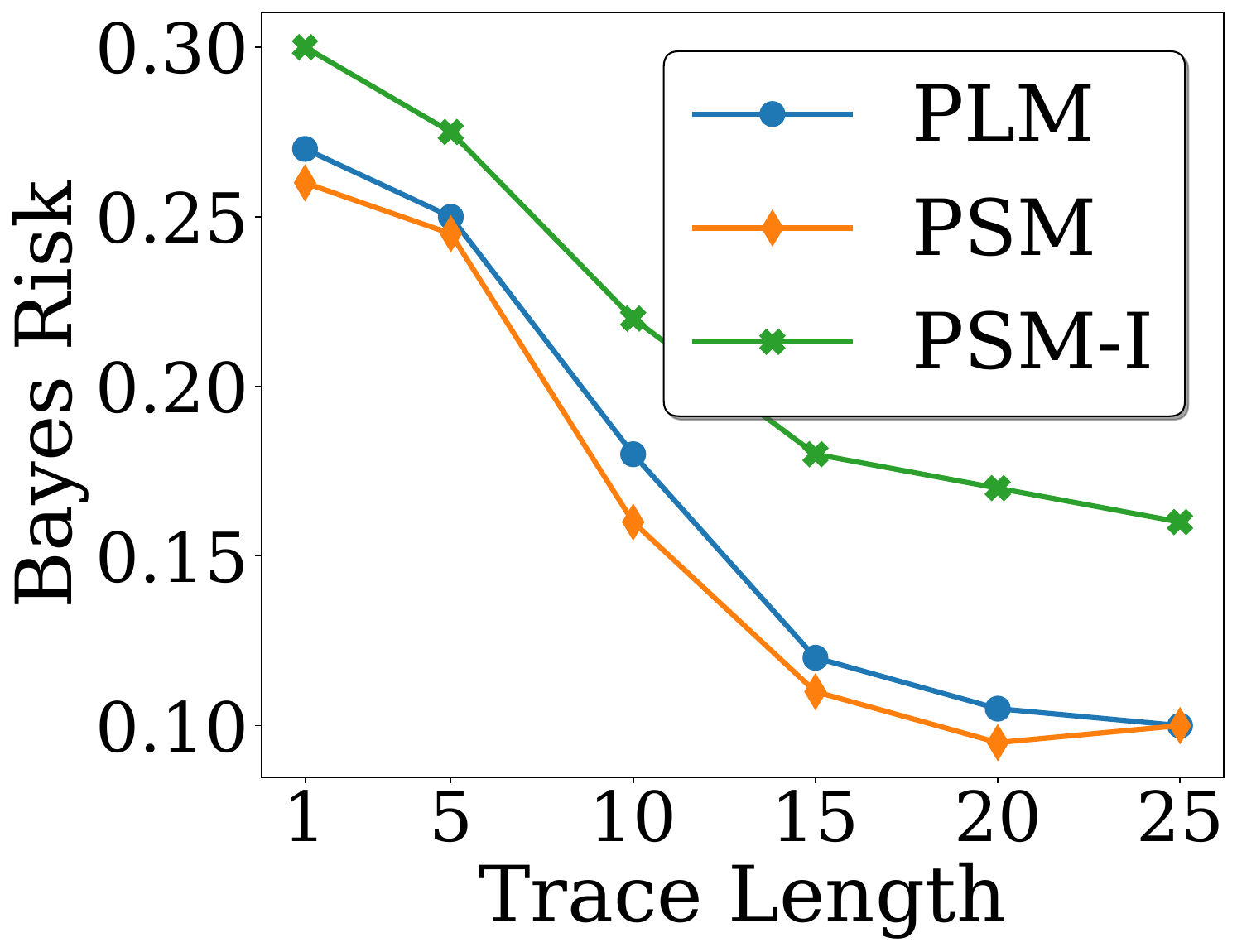} 
    \caption{Bayes risk for \emph{GeoTrace} at $\epsilon=0.1$.}
    \label{fig:geotrace_bayes}
\end{figure}

\begin{table}[t!]
\centering
\caption{Latency breakdown of client–server workflow.}
\label{tab:latency_breakdown}
\begin{tabular}{p{0.65\linewidth} r}
\toprule
\textbf{Operation} & \textbf{Duration (ms)} \\
\midrule
Device acquires location fix & 0.03 \\
\textbf{Apply PrivAR (PSM/PSM-I)} & \textbf{0.06} \\
Serialize JSON payload & 1.32 \\
Server-side object generation & 0.28 \\
Network round-trip (request + response) & 30.52 \\
Parse JSON response and render object & 1.70 \\
\midrule
Total End-to-End Latency & 33.91 \\
\bottomrule
\end{tabular}
\end{table}

\subsubsection{Evaluation} We describe the results below.

\paragraph{QoS in AR Gameplay} Figures~\ref{fig:catchable_objects} and~\ref{fig:Accumulated_loss} report AR gameplay QoS under dense and sparse virtual-object layouts. Across all privacy budgets, both PSM and PSM-I consistently outperform PLM. Under strong privacy ($\epsilon=0.1$), PSM-I preserves approximately 90--93\% object catchability in dense layouts and 75--80\% in sparse layouts, compared to substantially lower values for PLM. PSM achieves the highest catchability across all settings. In terms of accumulated loss (Fig.~\ref{fig:Accumulated_loss}), PSM-I reduces total object loss by approximately 30--40\% relative to PLM over longer traces, while PSM achieves reductions of up to 50\%. These gains are more pronounced in dense layouts, where small localization errors translate to larger gameplay penalties. Overall, PSM-I maintains strong AR utility while enabling substantially stronger trace-level privacy than per-update perturbation.


\begin{table}[!t]
\centering
\caption{Bayes risk under HMM-based trace inference for $\epsilon=0.1$. 
Higher Bayes risk indicates lower attacker confidence.} 
\label{tab:bayesrisk-hmm}
\begin{tabular}{c|ccc}
\toprule
Trace Length  & PLM & PSM & PSM-I \\
\midrule
1  & 0.327 & 0.326 & \textbf{0.335} \\
2  & 0.181 & 0.178 & \textbf{0.227} \\
3 & 0.134 & 0.139 & \textbf{0.167} \\
4 & 0.110 & 0.113 & \textbf{0.151} \\
5 & 0.096 & 0.098 & \textbf{0.144} \\
\bottomrule
\end{tabular}
\end{table}

\paragraph{Privacy}\label{sec:privacy} Figure~\ref{fig:geotrace_bayes} shows Bayes risk under a trace inference attack on the \emph{GeoTrace} dataset. PSM-I consistently achieves higher Bayes risk than both PLM and PSM across all trace lengths. At longer traces, PSM-I improves Bayes risk by up to 1.8$\times$, indicating significantly lower attacker confidence. Table~\ref{tab:bayesrisk-hmm} reports Bayes risk under a stronger HMM-based attacker that models user mobility as a first-order Markov process and performs Bayesian filtering over discretized states. As expected, Bayes risk decreases as the observation window length $L$ increases due to stronger temporal inference. While PLM and PSM exhibit similar trace-level leakage, PSM-I consistently maintains higher Bayes risk across all window lengths, demonstrating improved resistance even against mobility-aware inference.

\paragraph{Latency}  Table~\ref{tab:latency_breakdown} reports end-to-end latency for the full client–server pipeline. Privacy perturbation contributes only 0.06~ms per update, accounting for less than 0.2\% of the total latency (33.91~ms). The dominant overhead arises from network round-trips (30.52~ms) and JSON serialization and parsing (3~ms). Although total latency exceeds the 16~ms AR responsiveness threshold~\cite{deber2015much}, this overhead is unrelated to privacy. Crucially, PrivAR’s perturbation cost is negligible and does not limit real-time AR performance.





\subsection{Summary of Properties}\label{sec:keyprops}
\begin{itemize}
    \item \textbf{Robust Privacy.}  
  PSM and PSM-I provide GeoInd-style privacy, with PSM-I reducing trace-level inference risk.
    \item \textbf{High QoS.}  
    PSM achieves up to $2\times$ lower error than PLM; PSM-I preserves utility with a modest trade-off.
    \item \textbf{Low Latency Overhead.}  
    Privacy perturbation contributes less than 0.2\% of end-to-end latency.
    \item \textbf{Seamless Deployment.}  
    PrivAR is client-side and integrates into LB-AR pipelines without server-side changes.
\end{itemize}





\section{Related Work}\label{relatedworks}


\textbf{Location-Privacy Mechanisms.}
Cryptographic protocols~\cite{ShaoFine14} provide strong guarantees but incur latency and heavy mobile CPU use, unsuitable
for resource-constrained mobile devices. 
Spatial cloaking hides a user in a $k$‑anonymity region~\cite{LiClo07}, but lacks formal guarantees and is vulnerable to composition attacks~\cite{ganta2008composition}.  
Obfuscation with Bayesian adversary tuning~\cite{shokri2012protecting} maximizes privacy for a given prior, yet degrades sharply when the prior changes. LDP eliminates the trusted server but injects large noise~\cite{wang2022srr}, harming user‑centric QoS in mobile AR~\cite{athanasiou2025enhancing}. Geo‑Indistinguishability (GeoInd)~\cite{AndresGeo13} offers an elegant solution for location protection; its Planar Laplace Mechanism (PLM) perturbs each fix in $\mathcal{O}(1)$ time and is therefore the de facto baseline for mobile scenarios. Optimal GeoInd solutions~\cite{BordenabeOpt14,biswas2022privic,atmaca2024privacy} reduce error but require solving large linear programs or enumerating the output space, making them prohibitive for on-device LB-AR.  Our PSM inherits PLM’s closed‑form sampling efficiency while mitigating radial bias, and PSM-I further addresses GeoInd’s trace fragility.

\textbf{Trajectory‑Level Privacy.} 
Offline DP publication of whole traces~\cite{tian2017novel,cunningham2021real} yields strong guarantees but presupposes the entire trajectory, impractical for live AR.  Online adaptations exploit Markov models within a \emph{predefined} map~\cite{XiaoPro15,ZhangOnline19,cao2017quantifying}; unpredictable open‑world motion in LB‑AR breaks their assumptions.  Recent correlation attacks recover up to 93\% of traces~\cite{yu2023privacy} even after single-fix location privacy.

\textbf{Privacy in Mobile/AR Systems.} 
AR research has focused on access control prompts~\cite{jana2013enabling} and TEEs/isolated execution~\cite{costan2016intel,ratazzi2019pinpoint}.  These protect code and memory but not the live GPS stream itself, leaving apps to implement their own location privacy-preserving mechanism (LPPM).  To our knowledge, no prior work offers a practical, client‑side location‑privacy layer tailored to high‑frequency LB‑AR.

\section{Conclusion}\label{conclusion}

We present \emph{PrivAR}, the first fully client-side privacy framework for real-time LB-AR. PrivAR is deployable as a plug-and-play client component and requires no server-side changes. It introduces two mechanisms. PSM improves per-location QoS under GeoInd-style guarantees by reshaping the noise distribution with an exponentially decaying, staircase-shaped profile, reducing distortion compared to PLM. PSM-I extends PSM to high-frequency streams by maintaining a device-resident intermediate privacy buffer and applying selective reporting, reusing prior releases when movement is small to limit trace-level leakage. Across two public mobility datasets and an end-to-end Android LB-AR prototype (including our GeoTrace dataset), PrivAR achieves strong utility and privacy. PSM improves AR QoS by up to 50\% and reduces error by up to $2\times$ compared to PLM, without degrading empirical privacy under standard inference attacks. PSM-I further strengthens resistance to trajectory inference, improving Bayes risk by up to $1.8\times$. Both mechanisms incur negligible runtime overhead (less than 0.2\% of end-to-end latency), demonstrating that practical, real-time privacy protection for LB-AR is achievable without sacrificing responsiveness.

\section*{Acknowledgment}
We thank the reviewers and the revision editor for their insightful feedback and guidance, which significantly strengthened this paper. GPT-4 was used solely for language editing. This work was partially supported by the U.S. National Science Foundation under Grant CNS-2245689 (CRII), \#2426470 and \#2436428 and by a Meta Research Award in Privacy-Enhancing Technologies. 

\bibliographystyle{IEEEtran}

\bibliography{ref}

\end{document}